\documentclass[11pt,a4paper]{article}

\usepackage[english]{babel}
\usepackage{amsmath,amssymb,amsfonts,a4}
\usepackage{eucal,bbm,stmaryrd}
\usepackage{hyperref}
\usepackage{enumerate}
\usepackage{tikz}
\usetikzlibrary{patterns}

\begin{document}

\numberwithin{equation}{section}

\newtheorem{Thm}{Theorem}
\newtheorem{Def}{Definition}
\newtheorem{Lem}{Lemma}
\newtheorem{Rem}{Remark}
\newtheorem{Cor}{Corollary}

\newcommand{\Thmautorefname}{Theorem}
\newcommand{\Defautorefname}{Definition}
\newcommand{\Lemautorefname}{Lemma}
\newcommand{\Remautorefname}{Remark}
\newcommand{\Corautorefname}{Corollary}

\def\itemautorefname~#1\null{(T#1)\null}

\newcommand{\Pf}{\noindent{\bf Proof: }}
\newcommand{\qed}{\hspace*{3em} \hfill{$\square$}}

\newcommand{\R}{\mathbbm{R}}
\newcommand{\bR}{\overline{\R}}
\newcommand{\N}{\mathbbm{N}}

\newcommand{\E}{\mathbbm{E}}
\renewcommand{\P}{\mathbbm{P}}

\newcommand{\B}{\mathcal{B}}
\newcommand{\F}{\mathcal{F}}
\newcommand{\G}{\mathcal{G}}
\newcommand{\T}{\mathcal{T}}

\newcommand{\la}{\lambda}
\newcommand{\La}{\Lambda}
\newcommand{\si}{\sigma}
\newcommand{\al}{\alpha}
\newcommand{\be}{\beta}
\newcommand{\ep}{\epsilon}
\newcommand{\ga}{\gamma}
\newcommand{\de}{\delta}
\newcommand{\De}{\Delta}
\newcommand{\ph}{\varphi}
\renewcommand{\th}{\theta}
\newcommand{\vth}{\vartheta}
\newcommand{\ka}{\kappa}
\newcommand{\bka}{\bar{\ka}}
\newcommand{\bU}{\overline{U}}
\newcommand{\Lan}{\La_n}

\newcommand{\stm}{\setminus}
\newcommand{\lra}[1]{\stackrel{#1}{\longleftrightarrow}}
\newcommand{\Ra}{\Rightarrow}
\newcommand{\Lra}{\Leftrightarrow}

\newcommand{\sumn}{\sideset{}{^{\ne}}\sum}

\newcommand{\YY}{\mathcal{Y}}
\newcommand{\bY}{{\overline{Y}}}
\newcommand{\bYY}{\overline{\YY}}
\newcommand{\Ec}{\mathcal{E}}

\newcommand{\ee}{e_1}
\newcommand{\Gn}{\bYY_G}
\newcommand{\cTn}{\T_n}
\newcommand{\tn}{t_n^{\bY}}
\newcommand{\Tn}{T_n^{\bY}}
\newcommand{\ftn}{\mathfrak{t}_n}
\newcommand{\fm}{\mathfrak{m}}

\newcommand{\tcTn}{\tilde{\T}_n}
\newcommand{\ttn}{\tilde{t}_n^{\bY'}}
\newcommand{\tTn}{\tilde{T}_n^{\bY'}}
\renewcommand{\tt}{\tilde{t}}
\newcommand{\tT}{\tilde{T}}
\newcommand{\ttau}{\tilde{\tau}}
\newcommand{\tC}{\tilde{C}}
\newcommand{\tP}{\tilde{P}}
\newcommand{\tm}{\tilde{m}}

\thispagestyle{plain}
\title{A lower bound on the displacement of particles\\ 
in 2D Gibbsian particle systems}
\author{Michael Fiedler and Thomas Richthammer}
\maketitle
\begin{abstract}
While 2D Gibbsian particle systems might exhibit orientational order
resulting in a lattice-like structure, 
these particle systems do not exhibit positional order 
if the interaction between particles satisfies some weak assumptions.  
Here we investigate to which extent particles within a box of size $2n\times 2n$ 
may fluctuate from their ideal lattice position.  
We show that particles near the center of the box typically show a displacement 
at least of order $\sqrt{\log n}$. 
Thus we extend recent results on the hard disk model to particle systems with 
fairly arbitrary particle spins and interaction. 
Our result applies to models such as rather general continuum Potts type models,  
e.g. with Widom-Rowlinson or Lenard-Jones-type interaction. 
\end{abstract}

\section{Introduction}

At low temperature or at high particle density particles in two dimensions may arrange themselves into a lattice-like structure, which is usually referred to as solid or crystal. 
This behaviour can be observed in simulations 
(see e.g. E.P.~Bernard and W.~Krauth in \cite{BK} or S.C.~Kapfer and W.~Krauth in \cite{KK}), 
but could not yet be shown rigorously for any realistic model.
However, it can be shown that these lattice-like structures cannot be very rigid. 
For the 2D harmonic crystal R.~Peierls showed in \cite{P1, P2} that the mean  
displacement of a particle from its ideal lattice site in a box of size $2n\times 2n$ 
is of order $\sqrt{\log n}$. 
In particular this implies the absence of positional order (i.e. the conservation of translational symmetry). 
Building on ideas by N.D.~Mermin and H.~Wagner (\cite{MW, M}) concerning the absence of order in a more general context, 
the absence of positional order was first established for 2D Gibbsian particle systems by J.~Fr\"ohlich and C.-E.~Pfister (\cite{FP1,FP2}), 
and later these results were generalized by T.~Richthammer (\cite{R1,R2}). 
The behaviour of Peierl's harmonic crystal is believed to be typical for general 2D particle systems not just in terms of the absence of positional order 
but also in terms of the size of the fluctuations of particle positions. 
Indeed, recently T. Richthammer showed in \cite{R3} that this is true in the particular case of the hard disk model: 
Here the typical displacement of a particle near the center of a box of size $2n\times 2n$ 
is bounded from below by a constant times $\sqrt{\log n}$. 
The aim of this article is to generalize this result to fairly arbitrary 2D Gibbsian particle systems. 

The conditions we impose on the particle systems are similar to those imposed in \cite{R2} 
to guarantee the absence of positional order: 
Particles are described in terms of their positions in $\R^2$ and their internal properties 
(such as shape, orientation, magnetic spin, electric charge, particle type)
encoded in the particle spin.
The interaction is given by a two-particle potential function $U$, which will be assumed to satisfy the following conditions: 
\begin{itemize}
\item 
$U$ is symmetric and invariant under the translation of particles. 
\item 
$U$ is superstable and lower regular, so that the existence of Gibbs measures is guaranteed and we have Ruelle bounds on correlations. 
\item 
$U$ is allowed to have a singularity or hard core at the origin, 
but $U$ satisfies mild assumptions on the shape of the hard core and the type of convergence into the singularity or hard core.
\item 
Outside of the singularity/hard core $U$ can be decomposed into a sufficiently smooth part 
and a small part that only has to satisfy some integrability condition 
but does not have to satisfy any regularity conditions. 
Thus $U$ neither needs to be smooth nor continuous. 
\end{itemize} 
For details we refer to the next section.  
Some prominent examples of potential functions that are contained in our setting are 
the continuum Widom-Rowlinson model, Lennard-Jones type potentials, continuum Potts type models, 
the hard rod model and models of hard disks with random radii. 
Given an interaction of the above form and parameter values for the inverse temperature 
and the activity (i.e.\! the a-priori particle density) 
we consider the corresponding (grand canoncial) continuum Gibbs measures
as a model for the equilibrium states of the particle system at hand. 
We deal with the non-smooth part of the potential function $U$ 
by encoding it into an independent percolation process on edges between particles. 

In order to state and prove our result we use techniques from \cite{R2} and \cite{R3} 
(building on a method introduced in \cite{R1} and enhanced in \cite{MP}). 
Indeed, the formulation of our result is non-trivial, 
since there is no a-priori labelling of particles that would allow to pinpoint a specific particle and investigate its positional fluctuations. 
Instead we describe the fluctuations in terms of a transformation of particle-edge configurations, 
satisfying the following properties:
\begin{itemize}
\item 
All particles of a given configuration are shifted in a predefined direction.
Particles outside the box of size $2n\times 2n$ are not shifted at all and particles near the center of the box are shifted by an amount of order $\sqrt{\log n}$.
\item 
Local structures are almost preserved in that particles close to each other are shifted by almost the same amount and particles connected by an edge are shifted by the same amount. 
\item 
The probability measure describing the model is only mildly affected. 
\end{itemize}
We note that some of the properties above are in conflict, for instance it might happen that 
particles outside of the box and particles near its center are connected by an edge. 
It is therefore necessary to introduce a set of good particle-edge configurations, 
for which there are no conflicts. 
We give a fairly geometric description of this set
and show that it has probability close to 1. 

While the result presented here builds on the result on the special case in \cite{R3} and uses the same line 
of reasoning, we would like to point out some of the differences: 
\begin{itemize}
\item 
The edge process is a considerable complication that was not present in the special case of hard disks. 
\item 
The definition of the set of good configurations is more complicated due to the more general framework here. 
We thus have to put more effort into showing that good configurations are likely. 
\item 
All probability estimates are considerably harder and we have to deal with technical complications there. 
\item 
All probability estimates make use of Ruelle bounds, 
and thus have to be taken w.r.t.\! to the infinite volume Gibbs measures rather than the conditional Gibbs distributions in the finite volume $[-n,n]^2$. 
This implies that our result is not uniform in the boundary configurations. 
\end{itemize}
We also would like to point out that it should be possible to formulate our result 
in terms of a bound on the decay of positional correlation functions 
(such as the correlation functions used in the various computational investigations, see e.g. 
\cite{BK}, \cite{Eea} and references therein). 
However this is still work in progress and will be addressed in future work. 

The paper is structured as follows. 
In \autoref{Sec:2} we introduce the setting, 
specify the assumptions we make and state our main results. 
In \autoref{Sec:3} we give the main body of the proof. The proofs of the lemmas formulated in the previous sections are relegated to \autoref{Sec:4}.

\section{Setting and result} \label{Sec:2}

\subsection{Interactions} 

We consider the particle space $\R^2_S := \R^2 \times S$, 
where the spin space $(S,\F_S,\la_S)$ is a probability space such that 
$\{(\si,\si): \si \in S\} \in \F_S \otimes \F_S$.  
In most applications $S$ will be a metric space, 
e.g. a finite set or $\R^n$. 
The case $S=\{0\}$ corresponds to particles without spins.
On $\R^2$ we consider the Euclidean norm  $\|.\|$ and the supremum norm $\|.\|_\infty$, 
the Borel-$\si$-algebra $\B^2$ and the Lebesgue-measure~$\la^2$. 
On subsets of $\R^2$ we consider the corresponding restrictions. 
On product spaces such as $\R^2_S$ and $\R^2_{S^2}$ we consider the corresponding 
product-$\si$-algebras and product measures. 
We will usually denote particles, positions and spins by variants of the letters $y,x,\si$. 
We will abbreviate integration w.r.t. the above measures by $dy :=( \la^2 \otimes \la_S)(dy), 
dx := \la^2(dx), d\si := \la_S(d\si)$.  
With a little abuse of notation we will apply functions and operations on $\R^2$ to particles, 
e.g. we write $\|(x,\si)-(x',\si')\| := \|x-x'\|$ or $(x,\si) + x' := (x+x',\si)$
for $x,x' \in \R^2, \si,\si' \in S$, 
and similarly we will identify a set $\La \subset \R^2$ with $\La \times S \subset \R^2_S$. 
We call a function $f: \R^2_S \to \R$ continuous or differentiable, if 
$f(.,\si): \R^2 \to \R$ is continuous or differentiable for every $\si \in S$.

\begin{Def} A function $U \!: \R^2_{S^2} \to \bR := \R \cup \{\infty\}$ is called a (symmetric) 
potential if it is measurable and 
$U_{\si,\si'}(x) = U_{\si',\si}(-x)$ for all $x \in \R^2, \si,\si' \in S$. 
Its hard core is defined by $K^U := \{U = \infty\}$. 
\end{Def}
$U(x-x',\si,\si') = U_{\si,\si'}(x-x')$ represents the potential energy 
corresponding to two particles $(x,\si),(x',\si') \in \R^2_S$. 
The symmetry assumptions are such that $U$ is unchanged if the order of the particles is interchanged 
or if both particles are shifted by the same amount (i.e. $U$ is translation invariant). 
By abuse of notation we will consider a symmetric potential $U$ a function on $(\R^2_S)^2$ via 
$$
U((x,\si),(x',\si')) =  U_{\si,\si'}(x-x') \text{ for } x,x' \in \R^2, \si,\si' \in S.
$$
Similarly we will consider subsets of $\R^2_{S^2}$ as subsets of $(\R^2_S)^2$.

\begin{Def} \label{Def:enlargement}
A set $K \subset \R^2_{S^2}$ is called a symmetric set, if $1_K$ is a (symmetric) potential.  
The range of $K$ is defined by 
$$
\|K\| := \sup\{\|x\|: (x,\si,\si') \in K\}.
$$ 
For $\ep>0$ we define the $\ep$-enlargement of $K$ by 
$$
K_{\epsilon} := 
\{(x+z,\si,\si') \in \R^2_{S^2}: (x,\si,\si') \in K, \|z\| < \ep\}
$$
\end{Def}

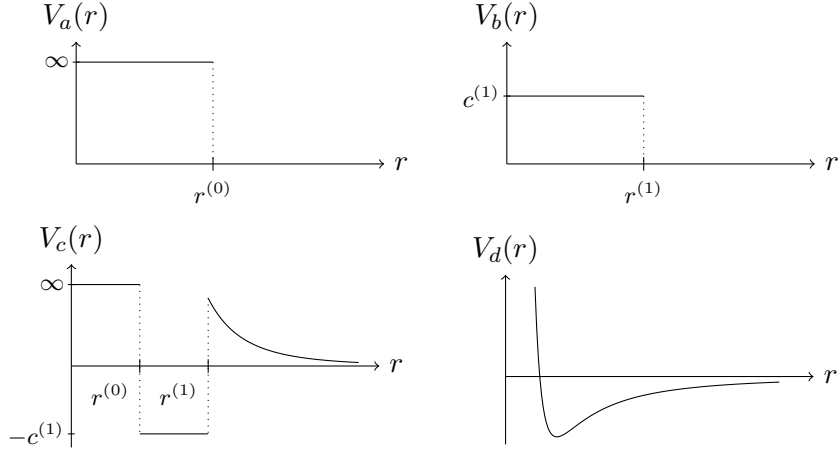
\begin{figure}[htb!] 
\centering
\begin{minipage}{0.42\textwidth}
\hspace*{0.2 cm} \begin{tikzpicture}[scale=0.9]
\draw[->] (0,0) -- (4.5,0) node[right] {$r$};
\draw[->] (0,0) -- (0,1.8) node[above] {$V_a(r)$};
\draw (-0.05,1.5) -- ( 0.05,1.5) node[anchor=east] {\footnotesize{$\infty$}}; 
\draw[domain=0:2] plot (\x,1.5) ;
\draw[dotted] (2,1.5)--(2,0);
\draw (2,0.1)--(2,-.1)node[anchor=north] {\footnotesize{$r^{(0)}$}};
\end{tikzpicture}
\end{minipage} \hspace{ 0.05 cm}
\begin{minipage}{0.42\textwidth}
\begin{tikzpicture}[scale=0.9]
\draw[->] (0,0) -- (4.5,0) node[right] {$r$};
\draw[->] (0,0) -- (0,1.8) node[above] {$V_b(r)$};
\draw (-0.05,1) -- ( 0.05,1) node[anchor=east] {\footnotesize{$c^{(1)}$}}; 
\draw[domain=0:2] plot (\x,1) ;
\draw[dotted] (2,1) -- (2,0);
\draw (2,0.1)--(2,-.1)node[anchor=north] {\footnotesize{$r^{(1)}$}};
\end{tikzpicture}
\end{minipage}

\begin{minipage}{0.42\textwidth}
\begin{tikzpicture}[scale=0.9]
\draw[->] (0,0) -- (4.5,0) node[right] {$r$};
\draw[->] (0,-1.2) -- (0,1.5) node[above] {$V_c(r)$};
\draw (-0.05,1.2) -- ( 0.05,1.2) node[anchor=east] {\footnotesize{$\infty$}}; 
\draw (-0.05,-1) -- ( 0.05,-1) node[anchor=east] {\footnotesize{$-c^{(1)}$}}; 
\draw[domain=0:1] plot (\x,1.2) ;
\draw[domain=1:2] plot (\x,-1) ;
\draw[samples=150, domain=2:4.2] plot (\x,{16/\x^4}) ;
\draw[dotted] (1,1.2)--(1,-1);
\draw[dotted] (2,1)--(2,-1);
\draw (1,0.1)--(1,-.1)node[anchor=north east] {\footnotesize{$r^{(0)}$}};
\draw (2,0.1)--(2,-.1)node[anchor=north east] {\footnotesize{$r^{(1)}$}};
\end{tikzpicture}
\end{minipage} \hspace{ 0.3 cm}
\begin{minipage}{0.42\textwidth}
\begin{tikzpicture}[scale=0.9]
\draw[->] (0,0) -- (4.5,0) node[right] {$r$};
\draw[->] (0,-1) -- (0,1.5) node[above] {$V_d(r)$};
\draw[samples=150, domain=0.43:4] plot (\x,{1.5/\x^2*(0.5/\x-1)}) ;
\end{tikzpicture}
\end{minipage}
\caption{Illustration of examples of $V$ in case of the absence of spins.}\label{fig:Potts}
\end{figure}

\begin{Def}\label{Def:smooth}
Let $\ee := (1,0)$.
A potential $U: \R^2_{S^2} \to \bR$ is called smoothly approximable 
if for every $\ga>0$ there exists a symmetric set $K \in \R^2_{S^2}$, 
and potentials $\bU,u: \R^2_{S^2} \to \R$ 
that decompose $U = \bU - u$ on $K^c$
such that the following conditions are satisfied: 
$K \supset K^U$ is sufficiently small in that  
\begin{align}
\begin{split} \label{equ:smoothK}
&\|K\| < \infty, \quad  \sup_{\si \in S}\int 1_{K \stm K^U}(x,\si,\si') dx d\si' <\ga,  \quad \\
&\forall \ep> 0: K_\ep \text{ is measurable}  \quad  \text{and} \quad 
\sup_{\si \in S} \int 1_{K_{\ep} \stm K}(x,\si,\si')dx d\si' \stackrel{\ep\to0}\longrightarrow 0,
\end{split}
\end{align}
$\bU$ is sufficiently smooth in that for some potential $\psi: \R^2_{S^2} \to \R$ 
\begin{align}
\begin{split}\label{equ:smoothU}
&\forall (x,\si,\si') \in K^c, t \in \R \text{ s.t. }
|t|\le \frac{\|x\|} 2,(x+t \ee,\si,\si')\in K^c: \\
&\hspace*{3 cm} |\partial_{\ee}^2\bU_{\si,\si'}(x+t\ee)| \le \psi_{\si,\si'}(x)\\
&\quad \text{and} \quad \sup_{\si \in S} \int  1_{K^c}(x,\si,\si') \psi_{\si,\si'}(x)(1 \vee \|x\|^2)dx d\si' <\infty,
\end{split}
\end{align}
and $u$ is sufficiently small in that 
\begin{align}
\begin{split} \label{equ:smoothu}
&\text{$u \ge 0$ on $K^c$}, \quad \sup_{\si \in S}\int 1_{K^c}(x,\si,\si')(u_{\si,\si'}(x) \wedge 1)dx d\si' <\ga \\
&\text{and}   \quad \sup_{\si \in S} \int 1_{K^c}(x,\si,\si')( u_{\si,\si'}(x) \wedge 1)\|x\|^2 dx d\si' <\infty.
\end{split}
\end{align}
\end{Def}
We note that in the above conditions, the potential values in $K$ are irrelevant. 
If potentials are considered as functions on $(\R^2_S)^2$, the above integrability conditions 
give bounds on the integrals w.r.t. the second particle that are uniform in the first particle. 
Smoothly approximable potentials may have a singularity or a hard core 
at the origin, and need not be smooth or even continuous.  
To show that the class of smoothly approximable potentials is very large, 
we give a class of examples. 
To keep the formulation simple, we will assume that the interaction 
only depends on the particle distance. 
\begin{Lem} \label{Lem:Potts} 
Let $N \ge 0$ and $r^{(0)},...,r^{(N)}$ be measurable symmetric functions on $S^2$ such 
that $0 \le r^{(0)} \le ... \le r^{(N)} = R$ for some $R > 0$. 
Suppose that $V: (0,\infty) \times S^2 \to \bR$ is measurable and symmetric such that  
\begin{itemize}
\item
$\{V = \infty\} = \{(r,\si,\si'): r < r^{(0)}_{\si,\si'}\}$ 
\item 
For every $\ep > 0$ there is an $M > 0$ such that $|V_{\si,\si'}(r)| \le M$ for all $\si,\si' \in S, r \in [r^{(0)}_{\si,\si'} + \ep, \infty)$.
\item 
For every $\ep > 0$ there is an $L > 0$ such that 
$V_{\si,\si'}$ is $L$-Lipschitz-continuous 
on $(r^{(i-1)}_{\si,\si'} + \ep,r^{(i)}_{\si,\si'}-\ep)$ for all $\si,\si' \in S^2$,
$i \in \{1,...,N\}$.
\item $V_{\si,\si'}$ is smooth on $(R,\infty)$ s.t. $|V_{\si,\si'}(r)|, |V_{\si,\si'}'(r)| \to 0$ 
for $r \to \infty$ and there are $c> 0,\al > 4$ such that 
$|V_{\si,\si'}''(r)| \le \frac{c}{r^{\al}}$ for all $r > R$, $\si,\si' \in S$.  
\end{itemize}
Then $U: \R^2_{S^2} \to \bR$ given by $U_{\si,\si'}(x):= V_{\si,\si'}(\|x\|)$ is 
smoothly approximable. 
\end{Lem}
For fixed spins $\si,\si'$ $V_{\si,\si'}$ may have a hard core 
and a finite number of discontinuities at the points $r^{(i)}_{\si,\si'}$. 
We have imposed a mild  regularity condition in between these points 
and a condition on the decay of $V''_{\si,\si'}$. 
Of particular interest may be the following types of interaction functions or variants of 
these: 
\begin{itemize}
\item[(a)] pure hard core repulsion $V_a(r) := \infty 1_{(0,r^{(0)})}(r)$, 
\item[(b)] pure soft core repulsion $V_b(r) := c^{(1)} 1_{(0,r^{(1)})}(r)$,
\item[(c)] well potential $V_c(r) := \infty 1_{(0,r^{(0)})}(r) - c^{(1)} 1_{(r^{(0)},r^{(1)})}(r) +  \frac {c^{(2)}} {r^3} 1_{(r^{(1)},\infty)}(r)$,
\item[(d)] Lenard Jones type potentials $V_d(r) := \frac{c^{(1)}}{r^{12}} - \frac{c^{(2)}}{r^6}$, 
\end{itemize}
where $c^{(i)} \ge 0$ are bounded measurable symmetric functions on $S^2$. 
For illustrations of these interaction functions see \autoref{fig:Potts}. 
Special cases of the pure hard core repulsion considered in case (a) are the Widom-Rowlinson model (see \cite{WR}) with $q \ge 2$ types of particles 
and particle radius $r > 0$, where we have $S=\{1,2\ldots,q\}$ 
and $r^{(0)}_{\si,\si'} = 2r 1_{\{\si \neq \si'\}}$, 
and a model of hard disks with random radii, where $S = (0,R)$ for some $R > 0$, 
and  $r^{(0)}_{\si,\si'} = \si + \si'$. 

An interesting example of an interaction that is not covered in the above lemma
is the hard rod model. 
Here we have  $S = S^1$ and particles $(x,\si)$ can be thought of as line segments 
$\{x + s \si: s \in [-r,r]\}$ for some given fixed $r > 0$. 
The interaction is given by a pure hard core repulsion $U = \infty 1_{K^U}$, where  
$$
K^U := \{(x,\si,\si') \in \R^2_{S^2}: \exists s,s' \in [-r,r]: x + s\si = s'\si'\}
$$ 
It is clear that $U$ is smoothly approximable (choosing $K = K^U, \psi =\bU = u = 0$).

\subsection{Gibbs measures}

A particle configuration is a locally finite subset of the particle space, 
i.e. a set $Y \subset \R^2_S$ such that $|Y \cap \La|<\infty$ for all bounded $\La \in \B^2$.
Particle configurations will usually be denoted by variants of the letter $Y$. 
We will write $Y_\La := Y \cap \La$ for the restriction of a configuration $Y$ to $\La \in \B^2$ and $YY' := Y \cup Y'$ for the concatenation of two configurations $Y$ and $Y'$.
Let $\YY$ denote the set of all particle configurations.  
Mainly we will use $\F_\YY := \si(N_B: B \in \B^2 \otimes \F_S)$ as $\si$-algebra on $\YY$, where $N_B(Y) := |Y \cap B|$. 
For $\La \in \B^2$ we also consider the $\si$-algebra
$\F_{\YY,\La} := \si(N_B: B \in \B^2(\La) \otimes \F_S)$. 
For any subset of configurations $\YY' \in \F_\YY$ we let $\F_{\YY'} := \F_\YY|_{\YY'}$. 
For a given bounded set $\La \in \B^2$ and boundary configuration 
$Y' \in \YY'$ we consider the probability measure $\nu_\La(.|Y')$
on $(\YY,\F_{\YY})$ that produces  a Poisson point process inside $\La$ 
and the deterministic configuration $Y'_{\La^c}$ outside $\La$, i.e. 
for every measurable function $f:\YY \to [0,\infty]$ we have 
\begin{align*}
\int f(Y )\nu_{\La}(dY|Y')
=e^{-\la^2(\La)}\sum_{k \ge 0}\frac1{k!} \int_{\La^k}
f(\{y_i:1\leq i\leq k\} \cup Y'_{\La^c})dy_1\dots dy_k.
\end{align*}
We note that $\nu_{\La}$ is a probability kernel from $(\YY,\F_{\YY,\La^c})$ to $(\YY,\F_{\YY})$. 

For $y,y' \in \R^2_S$ we write $yy' := \{y,y'\}$ for the edge between the two particles. 
In the following we will identify symmetric functions on pairs of particles with functions on edges between particles (by a slight abuse of notation). 
For a given potential $U$, for $Y,Y' \in \YY$ and  $\La \in \B^2$  we consider the energy sums 
\begin{align*}
H^U(Y) &:= \sum_{yy' \in E(Y)} U(y,y') \text{ for }E(Y) := \{yy': y \neq y' \in Y\},\\ W^U(Y,Y') &:= \sum_{yy' \in E(Y,Y')} U(y,y') \text{ for } 
E(Y,Y') = \{yy': y \in Y, Y' \in Y\} \text{ and } \\
H^U_\La(Y) &:= \sum_{yy' \in E_\La(Y)} U(y,y') \text{ for } E_\La(Y) := E(Y_\La) \cup E(Y_\La,Y_{\La^c}),
\end{align*}
whenever the sums have a well defined value in $\bR$. 
$H^U_\La$ is called the Hamiltonian w.r.t.~$U$ and $\La$. 
For given potential $U$, inverse temperature $\be > 0$, activity $z > 0$,
boundary configuration $Y' \in \YY$ and bounded $\La \in \B^2$ 
we define the corresponding conditional Gibbs distribution $\mu_\La^{U,\be,z}(.|Y')$ by 
\begin{align*}
\mu_{\La}^{U,\be,z}(dY|Y') &:= \frac1{Z_{\La}^{U,\be,z}(Y')}e^{-\be H_{\La}^U(Y)}z^{|Y_{\La}|}\nu_\La(dY|Y'), \text{ where } \\
Z_{\La}^{U,\be,z}(Y') &:= \int \exp(-\be H_{\La}^U(Y))z^{|Y_{\La}|}\nu_{\La}(dY|Y'). 
\end{align*}
It is interpreted as the equilibrium state of the particle system 
given by $U,\be,z$
for a fixed particle configuration $Y'$ outside of $\La$.  
We ensure that $\mu_\La^{U,\be,z}$ is well defined
by considering only appropriate boundary configurations: 
\begin{Def}\label{Def:admissible}
$\YY' \in \F_{\YY}$ is called an admissible set of boundary configurations 
w.r.t.~a potential $U$ and $\be,z> 0$ if for all $Y' \in \YY'$ 
and all bounded $\La \in \B^2$ 
$$
Y' \in \YY' \Leftrightarrow Y'_{\La^c} \in \YY', \;  \;
W^U(Y'_{\La},Y'_{\La^c}) \in \bR \text{ is well defined} \;\; \text{ and } \;\; Z_\La^{U,\be,z}(Y')<\infty.
$$
\end{Def}
For admissible $\YY'$ indeed the conditional Gibbs 
distributions w.r.t. every $Y' \in \YY'$ is well defined since 
$Z_{\La}^{U,\be,z}(Y') \ge \nu_\La(\{\emptyset\}|Y') > 0$ 
and since for every $Y \in \YY$, $Y' \in \YY'$ we have $Y_\La Y'_{\La^c} \in \YY'$, 
so $H^U_\La(Y_\La Y'_{\La^c}) = H^U(Y_\La) + W^U(Y_\La,Y'_{\La^c})$ is well defined.  
For admissible $\YY'$ $\mu_\La^{U,\be,z}$ gives a probability kernel 
from $(\YY',\F_{\YY',\La^c})$ to $(\YY,\F_{\YY})$. 
The overall equilibrium states of the particle system are now given by Gibbs measures, 
which are defined in terms of the so called DLR-condition: 

\begin{Def}\label{Def:Gibbs}
Let $U$ be a potential, $\be,z > 0$ and $\YY'$ admissible. 
A probability measure $\mu$ on $(\YY,\F_\YY)$ is called a corresponding Gibbs 
measure if $\mu(\YY') = 1$ and $\mu = \mu \otimes \mu_\La^{U,\be,z}$ for every bounded $\La \in \B^2$, 
i.e. for every measurable $f \!:\! \YY \to [0,\infty]$  
\begin{align*}
\int \mu(dY) f(Y)= \int \mu(dY') \int \mu_\La^{U,\be,z}(dY|Y') f(Y).
\end{align*}
The set of all corresponding Gibbs measures will be denoted by $\G_{\YY'}^{U,\be,z}$.
\end{Def}
We note that the DLR-condition is equivalent to the assertion that
the conditional distributions of $\mu$ w.r.t. $\F_{\YY,\La^c}$ are given 
by $\mu_\La^{U,\be,z}$. 
We also note that Gibbs measures respect the hard core of the 
underlying potential in that 
\begin{align} \label{equ:hcc}
\mu(\{Y \in\YY :\exists y \neq y'\in Y: (y,y') \in K^U\})=0,
\end{align}
which follows from the DLR-condition. A main tool for estimating expectations w.r.t. 
Gibbs measures will be Ruelle bounds: 
\begin{Def}\label{Def:Ruelle}
Let $U$ be a potential, $\be,z > 0$, $\YY'$ admissible and $\mu \in \G^{U,\be,z}_{\YY'}$. Let  
\begin{align*} 
\rho^{U,\be,z}_{\mu}(Y) := z^{|Y|} e^{-\be H^U(Y)}\int \mu(dY') e^{-\be W^U(Y,Y')}
\quad \text{ for finite $Y \in \YY$}. 
\end{align*}
$\rho^{U,\be,z}_{\mu}$ is called the correlation function for $\mu$.  
A constant $\xi \ge 0$ such that 
$$
\rho^{U,\be,z}_{\mu}(Y) \le \xi^{|Y|} \quad \text{ for every $\mu \in \G^{U,\be,z}_{\YY'}$  and every finite $Y \in \YY$}
$$
will be called a Ruelle bound for $U,\be,z,\YY'$. 
\end{Def}
\begin{Lem} \label{Lem:Ruellebound}
Let $U$ be a potential, $\be,z > 0$ and $\YY'$ admissible, 
and  let $\xi$ be a corresponding Ruelle bound. Then for every  
$\mu \in \G^{U,\be,z}_{\YY'}$ and every  
measurable function $f:(\R^2_S)^m\to [0,\infty]$ $(m \ge 0)$ we have 
\begin{equation} \label{equ:Ruellebound}
\int \mu(dY) \sumn_{y_1,\ldots,y_m \in Y} f(y_1,\ldots,y_m) \le 
\xi^m \int dy_1 \ldots \int dy_m f(y_1,\ldots,y_m).
\end{equation}
Here $\sumn$ denotes a sum over distinct particles. 
\end{Lem}
The following lemma gives important cases of potentials, 
for which a sensible set of admissible configurations can be given 
and Ruelle bounds exist. 
\begin{Lem} \label{Lem:ruelle}
Let $U$ be a potential and $\be,z > 0$.
\begin{itemize}
\item[(a)] If $U \ge 0$ then $\YY$ is admissible and $\xi := z$ 
is a Ruelle bound. 
\item[(b)] If $U$ is superstable and lower regular, then the set of tempered 
configurations $\YY_t$ is admissible and there is a Ruelle bound $\xi$. 
\end{itemize}
\end{Lem}
We note that (a) is trivial and for the proof of (b) (and the definition of superstability, 
lower regularity and temperedness) we refer to \cite{Ru}. 
Since we need Ruelle bounds, we have to impose further conditions on 
the class of potentials to be considered. 
While purely repulsive potentials work fine by the above lemma, 
for the other potentials considered in \autoref{Lem:Potts} we 
have to assume superstability and lower regularity. 
This requires sufficient repulsion at short distances 
and a suitable decay at large distances, see Proposition 1.4 of \cite{Ru}.

\subsection{Edge process}

In order to deal with the non-smooth part $u$ of the interaction we will 
introduce a suitable edge process. 
We define the space of particle-edge-configurations by
\begin{align*}
\bYY := \{(Y,B):Y \in \YY, B \subset E(Y)\},
\end{align*}
endowed with the $\si$-algebra $\F_{\bYY} := \si(N_A,N_{A_1,A_2}: A,A_1,A_2 \in \B^2\otimes \F_S)$, where $N_A(Y,B) := |Y \cap  A|$ and 
$N_{A_1,A_2}(Y,B) = |\{yy' \in B: y \in A_1,y' \in A_2\}|$.
To define a useful probability measure on edge configurations let $v \ge 0$ be a potential, $\La \in \B^2$ and $Y \in \YY$.
Let  $\bYY_{Y,\La} := \{(Y,B): B \subset E_\La(Y)\}$ and 
$\F_{\bYY_{Y,\La}} := \si(N_b: b \in E_\La(Y))$, where $N_b(Y,B) := 1_{\{b \in B\}}$. 
On $(\bYY_{Y,\La},\F_{\bYY_{Y,\La}})$ we consider the Bernoulli-measure $\pi^v_\La(.|Y)$,
such that 
\begin{align*}
&\text{$N_b, b \in E_\La(Y),$ are independent and  $\{0,1\}$-valued with}  \\
&\pi^v_\La(N_b = 1|Y) = 1 - e^{ -v(b)} \text{ for } b \in E_\La(Y). 
\end{align*}
Since $\F_{\bYY_{Y,\La}} = \F_{\bYY}|_{\bYY_{Y,\La}}$, 
we may consider $\pi^v_\La(.|Y)$ a probability measure on $(\bYY,\F_{\bYY})$. 
Indeed, $\pi^v_\La$ is a stochastic kernel from $(\YY,\F_{\YY})$ to $(\bYY,\F_{\bYY})$. 
By definition we have for every finite $B \subset E_\La(Y)$: 
\begin{equation}\label{equ:Bernoulli}
\pi^v_\La(\{(Y,B)\}|Y) = \prod_{b \in B} ( 1 - e^{ -v(b)}) \!\!\! \prod_{b \in E_{\La}(Y) \stm B} 
\!\!\! e^{ -v(b)}
= e^{- H^v_\La(Y)} \prod_{b \in B} (e^{v(b)}-1). 
\end{equation}
This is meant to motivate the introduction of the edge process:  
If we have a decomposition $U = \bU - u$ into a suitably nice potential $\bU$ 
and a suitably small potential $u \ge 0$, in  
$\mu^{U,\be,z}_{\La} \otimes \pi^{\be u}_\La$, the density terms involving $H^{\be u}_\La$ cancel
and we are only left to deal with $u$-terms on particle pairs connected by the edge process. 
For small $u$ there will very likely be only few edges to deal with. 
Finally, we note that in case of $u = 0$ (which is a sensible choice for some of the 
potentials we are interested in), $\pi_\La^{\be u}$ a.s. gives no edges, 
so we do not have to consider the edge process at all.

\subsection{Result}

Our aim is to investigate to which extent a particle system in equilibrium
 - given by a pair potential $U$, inverse temperature $\be > 0$ and 
activity $z> 0$ - 
typically deviates from a lattice structure in terms of positional order. 
We thus are interested in the fluctuations of the positions 
of particles near the center of a large bounded domain $\La \in \B^2$, 
when particles outside of $\La$ are kept fixed. 
For simplicity we only consider fluctuations in direction $\ee$  
and domains of the form $\La_n$: 
$$
\ee := (1,0) \quad \text{ and } \quad  \La_n = [-n,n]^2,  (n \in \{1,2,...\}). 
$$
For ease of notation we will often write dependencies on $\La_n$ as 
dependencies on $n$. 
The following theorem generalizes Theorem 1 of \cite{R3} to a setting 
similar to that of Theorem 2 of \cite{R2}.  
\begin{Thm}\label{Thm:1}
Let $U$ be a potential, $\be, z > 0$, $\YY' \in \F_{\YY}$ admissible and  
$\xi$ a corresponding Ruelle-bound. 
Suppose that $U$ is smoothly approximable, and let $K,\bU,u$ be a corresponding choice for $\ga := \frac 1 {3 \xi (1 \vee \be)}$. 
Furthermore let $\de >0$. 
There is a set $\Gn \in \F_{\bYY}$ and there are $C,N > 0$ 
such that for all $c \in [0,C]$ and $n \ge N$ there is a transformation $\cTn: \bYY \to \bYY$ 
with the following properties: 
\begin{enumerate}[\upshape(T1)]
\item  \label{itm:T1}
$\cTn$ is of the form $\cTn(Y,B) = (\{\Tn(y) \!:y\in Y\}, 
\{\Tn(y)\Tn(y')\! :yy'\in B\})$, 
where $\Tn(y) = y + \tn(y) \ee$  for some function $\tn: Y \to [0,\infty)$.
\item \label{itm:T2}
For all $\bY \in \bYY$: $\Tn(y) = y$ for $y \in \La_n^c$
 and $\Tn(y) \in \La_n$ for $y \in \La_n$. 
\item \label{itm:T3}
For all $\bY \in \Gn$: $\Tn(y)=y + c \sqrt{\log n} \ee$
for $y \in \La_{\sqrt n}$.
\item \label{itm:T4}
For all $\bY \! = \! (Y,B) \! \in \! \YY$, $y,y' \! \in \! Y$ \!: 
$\tn(y)=\tn(y')$ if $yy' \! \in  \! B$ or $(y,y') \! \in \! K$.
\item \label{itm:T5}
For all $\bY \in \YY$, $y,y' \in Y$: 
$|\tn(y) - \tn(y')| \le  \de \|y-y'\|$.
\item \label{itm:T6}
For every $\mu \in \G^{U,\be,z}_{\YY'}$ we have
 $\mu\otimes\pi_n^{\be u}(\Gn)\ge 1-\de$.
\item \label{itm:T7}
$\cTn$ and $\cTn^-$ are bijective and for every $\mu \! \in \! \G^{U,\be,z}_{\YY'}$ \!\! 
and every $D \!\in\! \F_{\bYY}$ 
we have  
$$\frac 1 2 (\mu\otimes\pi_n^{\be u}(\cTn D)
+ \mu\otimes\pi_n^{\be u}(\cTn^- D)) \ge \mu\otimes\pi_n^{\be u}(D) - \de, 
$$
where $\cTn^-$ is defined as $\cTn$ in (T1) with $\ee$ replaced by $-\ee$.
\end{enumerate}
\end{Thm}
\begin{figure}[htb!]
\begin{center}
\begin{tikzpicture}[scale=0.95]
 \draw (-2.5,-2.5) rectangle (2.5,2.5);
 \draw[dotted](- 0.8,-0.8) rectangle (0.8,0.8);
\foreach \x in {(-4,-1),(4,2),(1.7,0.3),(3.5,-0.6),
(-2.2,0.7),(-0.8,1.8),
(1.3,2),
(-1.2,-1.9), (-1,-1.6),
(0.9,-1.5), (0.5,-0.5),
(-0.2,0.5)}
\draw[fill=black] \x circle (2 pt);
\foreach \x in {
(-2.0,0.7),(-0.6,1.8),
(1.6,2),
(-0.9,-1.9), (-0.7,-1.6),
(1.4,-1.5), (1.0,-0.5),
(0.3,0.5)}
\draw \x circle (2 pt);
\draw (4,2)--(1.7,0.3); \draw (-2.2,0.7)--(-0.8,1.8); \draw (0.9,-1.5)-- (0.5,-0.5);
\draw[dashed] (-2,0.7)--(-0.6,1.8); \draw[dashed] (1.4,-1.5)-- (1.0,-0.5);
\node at (2.8,2.5) {$\Lambda_n$};
\node at (1.2,0.9) {$\Lambda_{\sqrt n}$};
\end{tikzpicture}
\end{center} 
\caption{Illustration of some of the properties of $\cTn$. 
The picture shows a configuration $\bY$ and its image $\cTn(\bY)$. 
$\bY$ is represented by black circles and drawn lines,
$\cTn(\bY)$ by white circles and dashed lines.}\label{fig:propertiestransformation}
\end{figure}
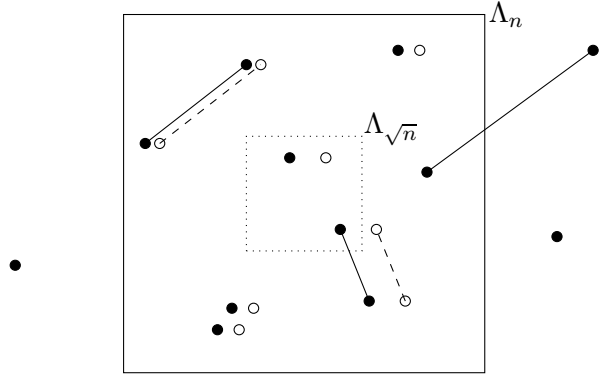
We are typically interested in choosing a small value for $\de > 0$. 
$\Gn$ is interpreted as a set of good configurations. 
We note that $\cTn$ also depends on the value of $c$, which we have not indicated in our notation. 
The properties in \autoref{Thm:1} can be interpreted as follows
(see Figure \ref{fig:propertiestransformation} for an illustration): 
By \autoref{itm:T1} $\cTn$ shifts every particle $y$ of a particle-edge configuration 
$\bY$ by an amount $\tn(y)$ in direction $\ee$, maintaining the edges between particles. The shift amount is allowed to depend on the full configuration. 
By \autoref{itm:T2} particles outside of $\La_n$ stay fixed, and particles within $\La_n$ remain in $\La_n$.  
By \autoref{itm:T3} particles inside of $\La_{\sqrt n}$ (i.e. particles near the origin) 
are shifted by an amount of order $\sqrt{\log n}$ if a good configuration is considered. 
By \autoref{itm:T4} particles almost at hard core distance and particles  
connected by an edge are shifted by the same amount.
By \autoref{itm:T5} the transformation almost preserves local structures: 
The Lipschitz property  implies that in a small region particles are shifted by almost the same amount; 
in particular if particles locally form a lattice-like structure, 
this structure is almost preserved by the transformation. 
By \autoref{itm:T6} good configurations have high probability. 
By \autoref{itm:T7} we have control over probabilities when applying the transformation.

We note that in \autoref{itm:T7} it is necessary to consider $\cTn^-$ along with $\cTn$; 
this is due to a Taylor-argument at the core of the estimate, 
where we need the first order terms to cancel (just as in Mermin-Wagner-type arguments). 
We note that it is also necessary to consider a set of good configurations, 
as otherwise some of the properties above are in conflict. 
Indeed, for example properties \autoref{itm:T2},\autoref{itm:T3} and \autoref{itm:T4} can not be satisfied simultaneously 
if $\bY$ contains a sequence of particles connecting 
$\La_{\sqrt n}$ to $\La_n^c$ such that 
the particles are too close to each other or connected by edges
(which would force them to be moved by the same distance). 
Such particle-edge configurations should be thought of as bad.
Finally we note that in \autoref{itm:T6} and \autoref{itm:T7}
we need to estimate probabilities w.r.t. $\mu$ rather than 
w.r.t. $\mu_n^{U,\be,u}(.|Y)$ for arbitrary $Y$, 
since the estimates using Ruelle bounds in finite volume may 
be non-uniform in boundary configurations

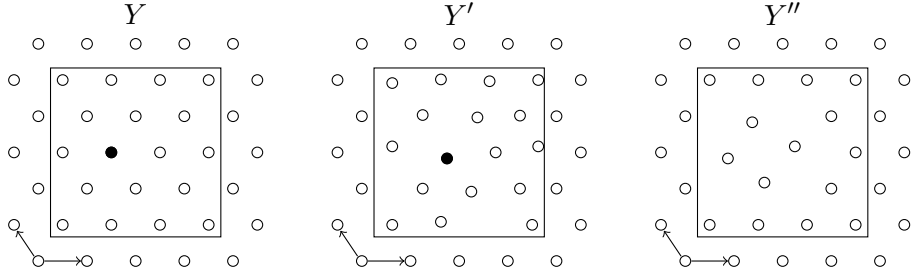
\begin{figure}[htb!]
\begin{center}
\begin{tikzpicture}[scale=0.8]
\node at (0,2.3) {$Y$};
\draw[->] (-1.5,-1.8)--(-0.9,-1.8);
\draw[->] (-1.67,-1.72)--(-1.95,-1.3);
\draw (-1.4,-1.4) rectangle (1.4,1.4); 
\foreach \x in {(-1.6,-1.8),(-0.8,-1.8),(0,-1.8),(0.8,-1.8),(1.6,-1.8),
(-2,-1.2),(-1.2,-1.2),(-0.4,-1.2),(0.4,-1.2),(1.2,-1.2),(2,-1.2),
(-1.6,-0.6),(-0.8,-0.6),(0,-0.6),(0.8,-0.6),(1.6,-0.6),
(-2,0),(-1.2,0),(-0.4,0),(0.4,0),(1.2,0),(2,0),
(-1.6,0.6),(-0.8,0.6),(0,0.6),(0.8,0.6),(1.6,0.6),
(-2,1.2),(-1.2,1.2),(-0.4,1.2),(0.4,1.2),(1.2,1.2),(2,1.2),
(-1.6,1.8),(-0.8,1.8),(0,1.8),(0.8,1.8),(1.6,1.8)}
\draw \x circle (2.5 pt); 
\draw[fill=black] (-0.4,0) circle (2.5 pt);
\end{tikzpicture}
\qquad 
\begin{tikzpicture}[scale=0.8]
\node at (0,2.3) {$Y'$};
\draw[->] (-1.5,-1.8)--(-0.9,-1.8);
\draw[->] (-1.67,-1.72)--(-1.95,-1.3);
\draw (-1.4,-1.4) rectangle (1.4,1.4); 
\foreach \x in {(-1.6,-1.8),(-0.8,-1.8),(0,-1.8),(0.8,-1.8),(1.6,-1.8),
(-2,-1.2),(-1.1,-1.2),(-0.3,-1.15),     (1.2,-1.2),(2,-1.2),
(-1.6,-0.6),(-0.6,-0.6),(0.2,-0.65),(1,-0.6),(1.6,-0.6),
(-2,0),(-1.1,0.1),(-0.2,-0.1),(0.6,0),(1.3,0.1),(2,0),
(-1.6,0.6),(-0.6,0.62),(0.3,0.58),(1,0.61),(1.6,0.6),
(-2,1.2),(-1.1,1.15),(-0.3,1.21),(0.5,1.18),(1.3,1.2),(2,1.2),
(-1.6,1.8),(-0.8,1.8),(0,1.8),(0.8,1.8),(1.6,1.8)}
\draw \x circle (2.5 pt); 
\draw[fill=black] (-0.2,-0.1) circle (2.5 pt);
\end{tikzpicture}
\qquad 
\begin{tikzpicture}[scale=0.8]
\node at (0,2.3) {$Y''$};
\draw[->] (-1.5,-1.8)--(-0.9,-1.8);
\draw[->] (-1.67,-1.72)--(-1.95,-1.3);
\draw (-1.4,-1.4) rectangle (1.4,1.4); 
\foreach \x in {(-1.6,-1.8),(-0.8,-1.8),(0,-1.8),(0.8,-1.8),(1.6,-1.8),
(-2,-1.2),(-1.2,-1.2),(-0.4,-1.2),(0.4,-1.2),(1.2,-1.2),(2,-1.2),
(-1.6,-0.6),           (0.8,-0.6),(1.6,-0.6),
(-2,0),                       (1.2,0),(2,0),
(-1.6,0.6),              (0.8,0.6),(1.6,0.6),
(-2,1.2),(-1.2,1.2),(-0.4,1.2),(0.4,1.2),(1.2,1.2),(2,1.2),
(-1.6,1.8),(-0.8,1.8),(0,1.8),(0.8,1.8),(1.6,1.8),
(-0.9,-0.1),(0.2,0.1), (-0.3,-0.5),(-0.5,0.5)}
\draw \x circle (2.5 pt); 
\end{tikzpicture}
\end{center}
\caption{Illustration of a particle identification procedure
identifying the particle $\ka$ 'in the centre' of $\La_n$. 
Above we consider three configurations such that $Y$ 
is a perfect lattice and $Y',Y''$ are perturbations of $Y$ 
such that $Y_{\La_n^c} \! = \! Y'_{\La_n^c} \! = \! Y''_{\La_n^c}$. 
We have indicated a lattice coordinate system and chosen the 
particle with lattice coordinate $(3,3)$ to represent the
central particle.   
$\ka(Y)$ (marked in black) corresponds to this choice.   
We set $\bka(Y) = \bka(Y') = \bka(Y'') = \ka(Y)$. 
$Y'$ still has a lattice like structure (even though particle (3,1) is missing) that allows us to identify 
$\ka(Y')$ (marked in black). In $Y''$ the lattice structure is broken 
in the vicinity of $(3,3)$, so we set $\ka(Y'') := \De$. 
}\label{fig:particlechoice}
\end{figure}

Since in our situation we have no a priori lattice structure, 
it is not clear how to refer to a specific particle and make statements about positional fluctuations or its displacement from its ideal lattice position. 
The above theorem is a substitute for such a statement; 
rather than focusing on a single particle the fluctuations are expressed in terms of a transformation on particle configurations preserving local structures without affecting probabilities in a significant manner. 
Indeed, if there is a procedure to pick a specific particle from a particle configuration 
that is not affected by local distortions of the configuration, 
the above result can be formulated in terms of an estimate 
of the displacement of the particle under consideration. 
\begin{Def} \label{Def:choice}
Measurable functions $\ka: \YY \to\R^2_S \cup\{\De\}$, $\bka: \YY \to \R^2 \cup \{\De\}$ 
are called a particle identification procedure in $\La_n$
if 
$$
\forall Y \in\YY: \ka(Y)\in Y_{\La_n} \cup\{\De\}, \bka(Y) \in \La_n \cup \{\De\} \quad 
\text{ and $\bka$ is $\F_{\YY,\La_n^c}$-measurable.}
$$  
$Y \in \YY$ is called $\ep$-stable w.r.t.\!\! $\ka$ for some $\ep > 0$ if for all distortions $Y' \in \YY$ of $Y$ 
of the form $Y' = \{y + \tau_y :  y \in Y\}$ with 
$\tau_y \in \R^2$  such that $Y'_{\La_n^c} = Y_{\La_n^c}$ and 
$ \|\tau_y - \tau_{y'}\| \le \ep \|y-y'\|$ for all $y,y' \in Y$ 
we have 
$$
\ka(Y') = \ka(Y) + \tau_{\ka(Y)} \quad (\text{and in particular }\ka(Y') \neq \De). 
$$
\end{Def}
We are mainly interested in situations, where $Y$ has a lattice-like structure. 
$\bka$ is supposed to identify the position of an ideal lattice site 
relative to a given configuration $Y_{\La_n^c}$, and  
$\ka(Y)$ is supposed to identify the actual particle of $Y$ corresponding to the ideal lattice site $\bka(Y)$. 
If $Y_{\La_n^c}$ does not have a sufficient lattice-like structure to produce a sensible value of $\bka(Y)$, one may set $\ka(Y) = \bka(Y) = \De$. 
If $\bka(Y) \neq \De$, but $Y_{\La_n}$ does not have a sufficiently lattice-like structure
to produce a sensible value of $\ka(Y)$, one may set $\ka(Y) = \De$. 
For an illustration see Figure \ref{fig:particlechoice}.
A configuration $Y$ is $\ep$-stable w.r.t.\!\! $\ka$ if the particle choice is not affected by distortions of $Y$, where particles are shifted in 
a way that leaves $Y_{\La_n^c}$ intact and the particle shifts satisfy 
an $\ep$-Lipschitz condition.

\begin{Cor} \label{Cor:1}
Let $U$ be a potential, $\be, z > 0$ and $\YY' \in \F_{\YY}$ admissible. 
Suppose that $U$ is smoothly approximable, and there is a Ruelle-bound. 
Let $\mu \in \G^{U,\be,z}_{\YY'}$.
Suppose that for every $n$, $\ka_n, \bka_n$ is a particle identification procedure in $\La_n$, 
such that $\bka_n \in \La_{\sqrt n /2} \cup\{\De\}$, 
and suppose that for some $\ep,\ep' > 0$ there are sets $\YY_n \in \F_{\YY}$ 
of $\ep$-stable configurations w.r.t. $\ka_n$ such that 
$\mu(\YY_n) \ge \ep'$ for every $n$.  
Then there are $\de,c,N > 0$ such that 
\begin{align*}
\forall n \ge N: \mu(\|\ka_n-\bka_n\|_\infty \ge c \sqrt{\log n}\})
\ge \de.
\end{align*}
\end{Cor}
Of course, it is still an open problem to show that for $\be \to \infty$ 
and/or $z \to \infty$ suitable particle systems exhibit a lattice-like structure, 
so it is not clear whether there is a suitable particle identification procedure such that for some small $\ep > 0$ $\ep$-stable configurations have a probability bounded away 
from $0$ (uniformly in $n$). 
The above corollary is merely meant to show that \autoref{Thm:1} can be considered a result 
on the order of magnitude of the typical particle displacement. 
Also the proof of the corollary shows how to use the transformation from \autoref{Thm:1}
to obtain properties of the particle system, and thus we hope that 
it contributes to a better understanding of \autoref{Thm:1}.

\section{Proof of the result} \label{Sec:3}

\subsection{Proof of \autoref{Thm:1}}

In this subsection we will give the main body of the proof of \autoref{Thm:1}. 
The proofs of the lemmas used here are relegated to the next section. 
Let $U$ be a potential, $\be,z > 0$, $\YY' \in \F_{\YY}$ admissible and 
$\xi$ a corresponding Ruelle-bound. 
Let $U$ be smoothly approximable, and choose $K,\bU,u,\psi$ accordingly for $\ga := \frac 1 {3 \xi(1 \vee \be)}$. 
With a view to \eqref{equ:smoothK} we may assume that $\{0\} \times S^2 \subset K$. 
Since \eqref{equ:smoothU} and \eqref{equ:smoothu} do not depend on what happens in $K$, we may also assume 
that $u = 0$ on $K$ and $\psi = 0$ on $K$. 
From now on we will consider potentials as translation invariant functions on $(\R^2_S)^2$ rather 
than functions on $\R^2_{S^2}$. 
We choose $\ep \in (0,1)$ such that $\sup \limits_{y \in \R^2_S} \int 1_{K_\ep \setminus K}(y,y') dy' < \ga$ and  consider the constants 
\begin{align}
\begin{split}\label{equ:constants}
c_K &:= \|K_\ep\| < \infty, \qquad 
c_\psi:= \xi \sup_{y \in \R^2_S} \int \psi(y,y')(\|y-y'\|^2 \vee 1)dy' < \infty, \\
c_u &:= \xi \sup_{y \in \R^2_S} \int (1_{K_\ep \setminus K^U}(y,y')+  \be u(y,y') \wedge 1) dy' < 1, \\
c_u' &:= \xi \sup_{y \in \R^2_S} \int  (1_{K_\ep \setminus K^U}(y,y') +  \be u(y,y') \wedge 1) \|y-y'\|^2 dy' <  \infty.
\end{split}
\end{align}
Here we have used properties  \eqref{equ:smoothK},\eqref{equ:smoothU},\eqref{equ:smoothu}
for the given choice of $\ga$. 
Let $\de > 0$. W.l.o.g. $\de$ may be assumed to 
be sufficiently small depending on the above objects and constants 
(to be specified whenever needed). 
Right away we will assume $\de < 10^{-6}$ and $\de < \frac 1 {c_K}$.  
Let $C := \de^2$, $N:= \frac 1 {\de^8}$ and 
let $c \in [0,C]$.
$U,\be,z,\YY',\xi,K,\bU,u,\psi,\ep,C,c,N$ will be fixed for the rest of the proof and we will drop dependencies on these objects from notations unless this leads to ambiguities. 
In the following we will consider various types of random objects. 
All of these objects can be seen to be measurable with respect to the corresponding 
$\si$-algebras (e.g. via Lemma 8 from \cite{R2}). 

\bigskip

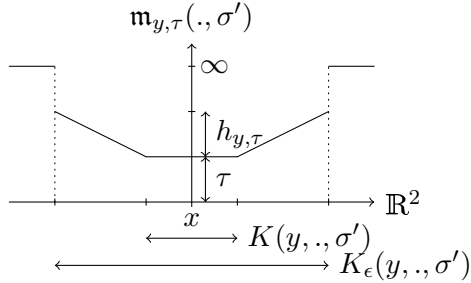
\begin{figure}[htb!]
\centering
\begin{tikzpicture}[line cap=round,line join=round,x=1cm,y=1cm,scale=0.6]
\draw[->] (-4,0.) -- (4,0.) node[right] {$\R^2$};
\foreach \x in {-3,-1,1,3}
\draw[shift={(\x,0)}] (0pt,2pt) -- (0pt,-2pt);
\draw[->] (0.,-.1) -- (0.,3.5) node[above] {$\fm_{y,\tau}(.,\si')$};
\foreach \y in {2,3}
\draw[shift={(0,\y)}] (2pt,0pt) -- (-2pt,0pt) ;
\draw (-1,1)--(1,1);
\draw (1,1)--(3,2);
\draw (-1,1)--(-3,2);
\draw (-4,3)--(-3,3);
\draw (4,3)--(3,3);
\draw[dotted](3,0)--(3,3);
\draw[dotted](-3,0)--(-3,3);
\node at (0,-0.4) {$x$};
\node at (.5,3) {$\infty$};
\draw[<->] (-1,-0.8)-- (1,-0.8) node[right]{$K(y,.,\si')$};
\draw[<->] (-3,-1.4)-- (3,-1.4) node[right]{$K_{\ep}(y,.,\si')$};
\draw[<->] (.3,0)--node[right]{$\tau$} (.3,1);
\draw[<->] (.3,1)--node[right]{$h_{y,\tau}$} (.3,2);
\end{tikzpicture}
\caption{Illustration of $\fm_{y,\tau}(.,\si')$, where $y = (x,\si)$, 
in case of $h_{y,\tau} \le \de \ep$.}
\label{fig:distortion}
\end{figure}

For fixed $n \ge N$ we consider the shift proposal $\ftn: [0,\infty) \to [0,\infty)$
\begin{align*}
\ftn(s) := \frac{3c}{\sqrt{\log n}}\log \frac n {n^{2/3} \vee (s \wedge n)}. 
\end{align*}
We note that  $\ftn(s) = c \sqrt{\log n}$ for $s \le n^{2/3}$ and $\ftn(s) = 0$ for $s \ge n$. 
$\ftn(\|.\|_\infty)$ serves as a first approximation of $t_n^{\bY}$. 
However, suppose that particle $y \in \R^2_S$ is shifted by an amount of $\tau \ge 0$, 
and we have another particle $y' \in \R^2_S$ close to $y$ 
for which $\ftn(\|.\|_\infty)$ proposes a shift amount that is too large. 
In view of \autoref{itm:T4} the shift of $y'$ has to be slowed down. 
We define a suitable slow-down function
$\fm_{y,\tau}: \R^2_S \to [0,\infty]$ by
\begin{align}
\begin{split}\label{Eq:1}
&\fm_{y,\tau}(y') := 
\begin{cases} 
\tau ,&\text{if }h_{y,\tau} > \de \ep\\
\tau +\frac{h_{y,\tau}}{\ep}d_K(y,y') +\infty 1_{(K_{\ep})^c}(y,y'),&\text{if }h_{y,\tau}\leq \de\ep,
\end{cases}\text{ where }\\  
&h_{y,\tau} := |\ftn(\|y\|_\infty-c_K)-\tau| \text{ and }   
 d_K(y,y') := \inf\{\ep' > 0: (y,y') \in K_{\ep'}\}.
\end{split}
\end{align}
$h_{y,\tau}$ represents the maximal possible shift amount of a particle at distance $c_K$ 
from $y$ proposed by $\ftn$, and $d_K(y,y')$ represents some notion of distance to $K$. 
We note that we have introduced the first case in the above definition to bound 
the slope of $\fm_{y,\tau}$ and we have introduced the $\infty$-part to restrict the 
influence of $\fm_{y,\tau}$ to a suitable neighbourhood of $y$. 
For an illustration of $\fm_{y,\tau}$ see \autoref{fig:distortion}. 

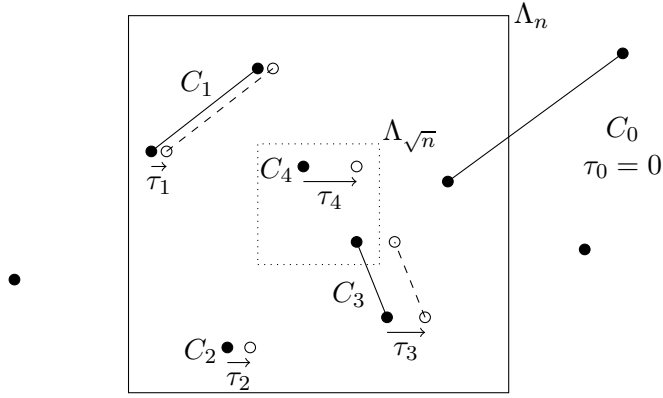
\begin{figure}[htb!]
\begin{center}
\begin{tikzpicture}[scale=1]
 \draw (-2.5,-2.5) rectangle (2.5,2.5);
 \draw[dotted](- 0.8,-0.8) rectangle (0.8,0.8);
\foreach \x in {(-4,-1),(4,2),(1.7,0.3),(3.5,-0.6),
(-2.2,0.7),(-0.8,1.8),
(-1.2,-1.9),
(0.9,-1.5), (0.5,-0.5),
(-0.2,0.5)}
\draw[fill=black] \x circle (2 pt);
\foreach \x in {
(-2.0,0.7),(-0.6,1.8),
(-0.9,-1.9),
(1.4,-1.5), (1.0,-0.5),
(0.5,0.5)}
\draw \x circle (2 pt);
\draw (4,2)--(1.7,0.3); \draw (-2.2,0.7)--(-0.8,1.8); \draw (0.9,-1.5)-- (0.5,-0.5);
\draw[dashed] (-2,0.7)--(-0.6,1.8); \draw[dashed] (1.4,-1.5)-- (1.0,-0.5);
\draw[->] (-2.2,0.5)-- node[below]{$\tau_1$}(-2,0.5);
\draw[->] (-1.2,-2.1)-- node[below]{$\tau_2$}(-0.9,-2.1);
\draw[->] (0.9,-1.7)-- node[below]{$\tau_3$}(1.4,-1.7); 
\draw[->] (-0.2,0.3)--node[below]{$\tau_4$} (0.5,0.3);
\node at (2.8,2.5) {$\Lambda_n$};
\node at (1.2,0.9) {$\Lambda_{\sqrt n}$};
\node at (4,1) {$C_0$};
\node at (4,0.5) {$\tau_0 = 0$};
\node at (-1.6,1.6) {$C_1$};
\node at (-1.55,-1.95) {$C_2$};
\node at (0.4,-1.2) {$C_3$};
\node at (-0.55,0.45) {$C_4$};
\end{tikzpicture}
\end{center}
\caption{Illustration of the construction of $\bY' := \cTn(\bY)$ by partitioning 
$\bY$ into clusters. $\bY$ is represented by black circles and drawn lines, 
$\bY'$ by white circles and dashed lines.}\label{fig:transformation}
\end{figure}

For a fixed particle-edge configuration $\bY = (Y,B) \in \bYY$ we will now construct the corresponding shift function $\tn$ and thus the transformation $\cTn$. 
We  proceed by recursively defining a partition $C_k$, $0 \le k \le m$,  
of $Y$, 
such that all particles of $C_k$ are shifted by the same amount $\tau_k$.
In view of \autoref{itm:T4} $C_k$ will be a $B$-cluster $C_B(Y')$ 
for a suitable $Y' \subset Y$, 
i.e. a set of the form 
\begin{align*}
C_{B}(Y') := \{y \in Y: \exists y_0,...,y_n \in Y: y_0 \in Y', y_n = y, 
y_{i-1}y_i \in B \forall i\}. 
\end{align*}
For an illustration of the definition of the transformation see \autoref{fig:transformation}.

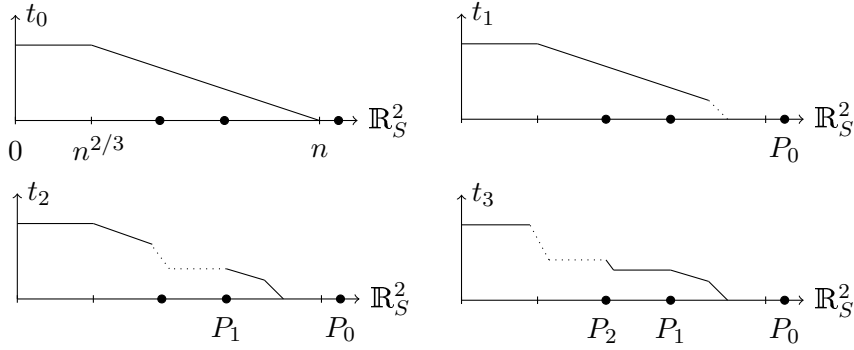
\begin{figure}[htb!]
\begin{minipage}{6 cm}
\quad 
\begin{tikzpicture}[scale=0.5]
\draw[->] (0,0) -- (9,0) node[right] {$\R^2_S$};
\draw[->] (0,0) -- (0,2.8) node[right] {$t_0$};
\draw (0,0.1) -- (0,-0.1);
\node at (0,-0.8) {$0$};
\draw (2,0.1) -- (2,-0.1); 
\node at (2.2,-0.7) {$n^{2/3}$};
\draw (8,0.1) -- (8,-0.1);
\node at (8,-0.8) {$n$};
\draw (0,2)--(2,2);
\draw (2,2)--(8,0);
\foreach \x in {8.5,5.5,3.8} \draw[fill=black] (\x,0) circle (3pt);
\end{tikzpicture}

\hspace*{0.5 cm}
\begin{tikzpicture}[scale=0.5]
\draw[->] (0,0) -- (9,0) node[right] {$\R^2_S$};
\draw[->] (0,0) -- (0,2.8) node[right] {$t_2$};
\draw (0,0.1) -- (0,-0.1);
\draw (2,0.1) -- (2,-0.1); 
\draw (8,0.1) -- (8,-0.1);
\draw (0,2)--(2,2);
\draw (2,2) -- (3.55,1.45);
\draw[dotted] (4,0.8)-- (3.5,1.47);
\draw[dotted] (5.5,0.8)--(4,0.8);
\draw (5.5,0.8)-- (6.5,0.5);
\draw (7,0)--(6.5,0.5);
\foreach \x in {8.5,5.5,3.8} \draw[fill=black] (\x,0) circle (3pt);
\node at (8.5,-0.8) {$P_0$};
\node at (5.5,-0.8) {$P_1$};
\end{tikzpicture}
\end{minipage}
\;\;
\begin{minipage}{6 cm}
\begin{tikzpicture}[scale=0.5]
\draw[->] (0,0) -- (9,0) node[right] {$\R^2_S$};
\draw[->] (0,0) -- (0,2.8) node[right] {$t_1$};
\draw (0,0.1) -- (0,-0.1);
\draw (2,0.1) -- (2,-0.1); 
\draw (8,0.1) -- (8,-0.1);
\draw (0,2)--(2,2);
\draw (2,2)--(6.5,0.5);
\foreach \x in {8.5,5.5,3.8} \draw[fill=black] (\x,0) circle (3pt);
\draw[dotted] (7,0)--(6.5,0.5);
\node at (8.5,-0.8) {$P_0$};
\end{tikzpicture}

\begin{tikzpicture}[scale=0.5]
\draw[->] (0,0) -- (9,0) node[right] {$\R^2_S$};
\draw[->] (0,0) -- (0,2.8) node[right] {$t_3$};
\draw (0,0.1) -- (0,-0.1);
\draw (2,0.1) -- (2,-0.1); 
\draw (8,0.1) -- (8,-0.1);
\draw (0,2)--(1.8,2);
\draw[dotted] (2.3,1.07)-- (1.8,2);
\draw[dotted] (2.3,1.07)--(3.8,1.07);
\draw (3.8,1.07) -- (4,0.8);
\draw (5.5,0.8)--(4,0.8);
\draw (5.5,0.8)-- (6.5,0.5);
\draw (7,0)--(6.5,0.5);
\foreach \x in {8.5,5.5,3.8} \draw[fill=black] (\x,0) circle (3pt);
\node at (8.5,-0.8) {$P_0$};
\node at (5.5,-0.8) {$P_1$};
\node at (3.7,-0.8) {$P_2$};
\end{tikzpicture}
\end{minipage}
\caption{1D qualitative illustration of the shift profiles, 
supposing every cluster $C_k$ consists of a single point $P_k$ 
and $h_{y,\tau} \le \de \ep$ for all distortions. 
$0$ represents the center of $\La_n$ and $n$ the boundary. 
The height of the shift profile $t_0$ is given by $c \sqrt{\log n}$. 
The distortion due to the particle under consideration is marked by a dotted line. 
While particle $P_0$ does not have an impact on the following particles, 
particle $P_1$ slows down the shift of $P_2$, so that the distortion due to particle $P_2$ is even more pronounced.  
}\label{fig:shift}
\end{figure}

In the recursive step we use the information on the shift amount of the particles 
considered so far and slow down $\ftn(\|.\|_\infty)$ accordingly 
to get a modified shift profile $t_{k}$. 
This shift profile will be used to determine the next set of points 
and their respective shift amounts. 
For a formal definition of this recursive scheme 
let $t_0 := \ftn(\|.\|_\infty)$, $C_0 := C_B(Y_{\La_n^c})$ and $\tau_0 := 0$. 
For the recursive step let $k \ge 1$ and define $t_k: \R^2_S \to [0,\infty)$ by 
\begin{align*}
t_k := t_0 \wedge \bigwedge_{l < k} \bigwedge_{y \in C_l} \fm_{y,\tau_l}.
\end{align*}
Let $P_k$ be the set of points of $Y \stm \bigcup_{l=0}^{k-1} C_l$ at which $t_k$ is minimal, 
let $\tau_k$ be the corresponding minimal value and $C_k = C_B(P_k)$. 
This finishes the recursive step. 
Let $m$ be the smallest value such that $\bigcup_{l=0}^m C_l = Y$ (so that 
the recursion stops after defining $t_{m+1}$ since it is not possible to choose $P_{m+1}$). 
If in the construction at some point we have a $y \in C_k$ s.t. $h_{y,\tau_k} > \de \ep$ 
and thus $\fm_{y,\tau_k} = \tau_k$, and $k$ is minimal in that respect, we set $m^* := k$. 
Otherwise we set $m^* := m$. 
For an illustration of the sequence of shift profiles used in the construction 
see \autoref{fig:shift}. 
Now we define 
$$
\tn(y) := \tau_k \quad \text{  for } y \in C_k
$$
and $\Tn$ and $\cTn$ according to \autoref{itm:T1}. 
It remains to show that this construction satisfies \autoref{itm:T2} - \autoref{itm:T7}. Properties \autoref{itm:T2}, \autoref{itm:T4}, \autoref{itm:T5}
are almost immediate form the construction: 
\begin{Lem}\label{Lem:transproperties}
For every $\bY =(Y,B) \in \YY$ and all $y,y' \in Y$ we have  
\begin{align} 
& 
y \in \La_n^c \Ra \Tn(y) = y \quad \text{ and } \quad y \in \La_n \Ra \Tn(y) \in \La_n, 
\label{equ:transLan}\\
&(y,y') \in K \text{ or } yy' \in B \quad \Ra \quad \tn(y) = \tn(y'),
\label{equ:transconstant}\\
&|\tn(y) - \tn(y')| \le \de \|y-y'\|. 
\label{equ:transLip}
\end{align}
\end{Lem}
With a view to property \autoref{itm:T3} we now introduce the set of good configurations:
\begin{equation*}
\begin{split}
&\Gn := \{(Y,B) \in \bYY: \forall y \in Y, y' \in C_{B_+}(y): 
\|y'\|_\infty \le 
\frac 1 \de (1 \vee \|y\|_\infty \log \|y\|_\infty)
\},\\
&\hspace*{3 cm} \text{ where }B_+ := B \cup \{yy': (y,y') \in K_\ep\}.
\end{split}
\end{equation*} 
Thus a configuration is good, if for every particle its $B_+$-cluster does not 
extend too far outwards (depending on the position of the particle).  
Property \autoref{itm:T3} of the transformation is implied by the following lemma:  
\begin{Lem} \label{Lem:goodprop}
For $\bY = (Y,B) \in \Gn$ we have $m^* = m$ and for all $y \in Y$: 
$$
y \in \La_{\sqrt n} \quad \Ra \quad \tn(y) = c \sqrt{\log n}.
$$
\end{Lem}
Before turning to probability estimates we study the effect of the transformation 
on probability measures. It can be described in terms of a certain density function. 
To formulate this property we first need to establish bijectivity: 

\begin{Lem}\label{Lem:bijective}
The transformations $\cTn,\cTn^-: \bYY \to \bYY$ are bijective. 
\end{Lem}
Here $\cTn^-$ is defined as $\cTn$ with $\ee$ replaced by $-\ee$, 
i.e. the two transformatoins shift particles by the same amount, 
but in the opposite direction. 
\begin{Lem} \label{Lem:density}
Let $\mu \in \G$, then $(\mu \otimes \pi_n) \circ \cTn$ is absolutely continuous w.r.t. 
$\mu \otimes \pi_n$ with density $\ph_n$, i.e. for every  measurable function $f: \bYY \to [0,\infty)$ we have 
\begin{equation} \label{equ:density}
\int \mu \otimes \pi_n(d \bY) \ph_n(\bY)  f(\cTn(\bY))  = 
\int \mu \otimes \pi_n(d \bY) f(\bY),
\end{equation}
where the density is given by 
\begin{equation} \label{equ:defdensity}
\ph_n(\bY)=\frac{e^{-\be H_{\La_n}^{\bU}(\cTn(\bY))}}{e^{-\be H_{\La_n}^{\bU}(\bY)}}
\th_n(\bY) 
\text{ with } \th_n(\bY) = \prod_{k=1}^{m}  \prod_{y \in P_k} \left|1+\partial_{\ee} t_k(y)\right|.
\end{equation}
Here $H^{\bU}_{\La_n}(\bY) := H^{\bU}_{\La_n}(Y)$ and
$P_k, t_k, m$ are the objects from the construction of $\cTn(\bY)$. 
$\cTn^-$ has a corresponding property with density $\ph_n^-$. 
\end{Lem} 
The proof of the preceding lemma will also show that $\ph_n$ is $\mu \otimes \pi_n$-a.s. 
well defined in that the derivatives exist. 
We now turn to the remaining probability estimates. 
Property \autoref{itm:T6} is shown in the following lemma: 
\begin{Lem} \label{Lem:goodprob}
For every $\mu \in \G$ we have $\mu \otimes \pi_n(\Gn^c) \le \de$. 
\end{Lem}
For property \autoref{itm:T7} we first show the
following energy and density estimates: 
\begin{Lem}\label{Lem:energyestimate}
For every $\mu \in \G$ we have $\be \int S_1 d\mu \otimes \pi_n \le \de$, 
where  
$$
S_1(\bY) :=  |H^{\bU}_{\La_n}(\cTn(\bY)) + H^{\bU}_{\La_n}(\cTn^-(\bY)) - 2 H^{\bU}_{\La_n}(\bY)|.
$$
\end{Lem}
\begin{Lem}\label{Lem:densityestimate}
For every $\mu \in \G$ we have $\int S_2 d\mu \otimes \pi_n \le \de$, 
where  
$$
S_2(\bY) := |\log (\th_n(\bY) \th_n^-(\bY))|.
$$
\end{Lem}
So for arbitrary $\mu \in \G$ and $D \in \F_{\bYY}$ we have  
\begin{align*}
&\mu \otimes \pi_n (D) - \frac {\mu \otimes \pi_n(\cTn D) + \mu \otimes \pi_n(\cTn^-D)} 2   
= \int   1_D (1 - \frac{\ph_n +\ph_n^-} 2) d\mu \otimes \pi_n\\
&\le \int   1_D (1 - (\ph_n \ph_n^-)^{\frac 1 2}) d\mu \otimes \pi_n
\le \int \frac 1 2 |\log (\ph_n \ph_n^-)| d\mu \otimes \pi_n \le \de.
\end{align*}
In the first step we have used the bijectivity of $\cTn, \cTn^-$ and \eqref{equ:density} and the correspoding result for $\cTn^-$. 
In the second step we have estimated the arithmetic with the geometric mean, and in the third step we have used that 
$1-x \le - \log x \le |\log x|$ for all $x \ge 0$ and $1_D \le 1$. 
In the last step we have used the definition of $\ph_n$ and $\ph_n^-$ and Lemmas \ref{Lem:energyestimate} and \ref{Lem:densityestimate}. 
Thit establishes property \autoref{itm:T7} 
and completes the proof of the theorem. \qed

\subsection{Proof of \autoref{Cor:1}}

Let $U$ be a potential, $\be,z> 0$, $\YY'$ admissible, such that there is a Ruelle bound. 
Let $U$ be smoothly approximable and choose $K,\bU,u$ accordingly for 
$\ga := \frac 1 {3 \xi (\be \vee 1)}$. 
Let $\mu \in \G_{\YY'}^{U,\be,z}$.
For every $n$ let $\ka_n,\bka_n$ be a particle identification procedure 
in $\La_n$ such that $\bka_n \in \La_{\sqrt n/2} \cup \{\De\}$, and 
for $\ep,\ep' > 0$ let $\YY_n \in \F_\YY$ a set of $\ep$-stable configurations w.r.t. $\ka_n$ such that $\mu(\YY_n) \ge \ep'$ for every $n$. 
For $\de := \ep \wedge \frac{\ep'} 4$ choose $\Gn \in \F_{\bYY}$ and $C,N > 0$ 
according to \autoref{Thm:1} and let $n \ge N$. W.l.o.g. $C \le \frac 1 2$ so that 
$$
\sqrt n /2 + C \sqrt{\log n} < \sqrt n.
$$ 
Let $k \ge 1$ such that $\frac 1 {2k} \le \de$, $c := \frac C k$ and let 
$$
D_n^{(i)} = \{\|\ka_n-\bka_n + ic \sqrt{\log n}\ee\|_\infty < \frac {c} {2} \sqrt{\log n}\} \text{ for } i \in \{-k,...,k\}.
$$
For the given values of $n$ and $c$ let $\cTn: \bYY \to \bYY$ be a transformation satisfying properties \autoref{itm:T1} - \autoref{itm:T7}. We claim that  
\begin{equation} \label{equ:transD}
\T_n (D_n^{(0)} \cap \Gn \cap \YY_n) \subset D_n^{(-1)}.
\end{equation}
For the proof let $\bY = (Y,B) \in D_n^{(0)} \cap \Gn \cap \YY_n$ 
and $\bY' = (Y',B') := \cTn(\bY)$. 
$Y \in \YY_n$ implies that $\ka_n(Y') = \ka_n(Y) + \tn(\ka_n(Y))\ee$, 
using $\de \le \ep$ and properties \autoref{itm:T2} and \autoref{itm:T5}. 
Since $Y \in D_n^{(0)}$, $\bka_n(Y) \in  \La_{\sqrt n/2}$ and $\sqrt n /2 + \frac c 2 \sqrt{\log n} < \sqrt n$, we have $\ka_n(Y) \in \La_{\sqrt n}$. 
Because of $\bY \in \Gn$ and \autoref{itm:T3} we have $\tn (y) = c \sqrt{\log n}$ for all $y \in \La_{\sqrt n}$, and in particular for $y = \ka_n(Y)$.   
Thus the above implies 
$$
\ka_n(Y') = \ka_n(Y) + c \sqrt{\log n} \ee.
$$ 
On the other hand we have $Y'_{\La_n^c} = Y_{\La_n^c}$ by \autoref{itm:T2} and 
by the $\F_{\YY,\La_n^c}$-measurability of $\bka_n$ this implies 
$$
\bka_n(Y') = \bka_n(Y).
$$ 
Since $Y \in D_n^{(0)}$ the above implies that $Y' \in  D_n^{(-1)}$, which 
completes the proof of \eqref{equ:transD}. 
Similarly we have $\T_n^- (D_n^{(0)} \cap \Gn \cap \YY_n) \subset D_n^{(1)}$. 
Since the events $D_n^{(i)}$ do not depend on edge configurations at all, 
\begin{align*}
&\frac 1 2(\mu(D_n^{(1)}) + \mu(D_n^{(-1)})) \ge \mu \otimes \pi_n(D_n^{(0)} \cap \Gn \cap \YY_n) - \de \\
&\ge \mu(D_n^{(0)} \cap \YY_n) - \mu \otimes \pi_n(\Gn^c) - \de
\ge \mu(D_n^{(0)} \cap \YY_n)  - 2 \de,  
\end{align*}
where we have also used \autoref{itm:T7} and \autoref{itm:T6}.
We obtain similar estimates for $i \in \{2,...,k\}$ by considering 
transformations corresponding to the translation distance $i c \le C$ instead of $c$. 
Adding these inequalities and using that the $D_n^{(i)}$ are disjoint we get 
$$
\frac 1 2 \ge k (\mu(D_n^{(0)} \cap \YY_n)  - 2 \de), \quad \text{ i.e. }
\mu(D_n^{(0)} \cap \YY_n) \le 3 \de,
$$
which implies the conclusion of the corollary, since 
\begin{align*}
\quad \mu\Big(\|\ka_n - \bka_n\|_\infty \ge \frac{c} 2 \sqrt{\log n}\Big) 
\ge \mu(\YY_n) - \mu(D_n^{(0)} \cap \YY_n) \ge \ep' - 3 \de \ge \de. \!\!\!\!\! 
\text{\qed}
\end{align*}

\section{Proof of the lemmas} \label{Sec:4}

\subsection{Potts-type models: \autoref{Lem:Potts}}

Let $U$ be a potential of the form given in the formulation of the Lemma. 
W.l.o.g. $R \ge 1$. 
Let $\ga > 0$. Let $K := K^U_{\ep_0} = \{(x,\si,\si'): \|x\| < r^{(0)}_{\si,\si'} + \ep_0\}$ for 
sufficiently small $\ep_0 \in (0,1)$, so that 
$$
\int 1_{K \stm K^U}(x,\si,\si') dx = 2 r^{(0)}_{\si,\si'} \ep_0\pi + \ep_0^2\pi
\le 2R\ep_0\pi + \ep_0^2\pi  < \ga. 
$$
$K$ is easily seen to satisfy \eqref{equ:smoothK}. 
Next we want to approximate $V$ by a smooth function. 
Let $\ep \in (0,\frac {\ep_0} 2)$ such that $N\pi 8R\ep < \frac 1 2 \ga$. 
Choose $M,L > 0$ w.r.t. this $\ep$ according to the boundedness and Lipschitz condition on $V$. 
Let  $\de\in (0,\frac \ep 2)$ such that $2 \de L (R+\ep)^2\pi < \frac 1 2 \ga$.
Let $\hat V_{\si,\si'}(r) := M$ if $r \in [r^{(i)}_{\si,\si'}- \ep-\de, r^{(i)}_{\si,\si'} + \ep+\de]$ for some $i$ 
or $r \le r^{(0)}_{\si,\si'}$ or $r \ge R$ and $\hat V_{\si,\si'}(r) := V_{\si,\si'}(r) + \de L$ otherwise. 
Let $f_{\de}:\R \to [0,\infty)$ be a sufficiently smooth symmetric probability density 
with support in $(-\de,\de)$, and use $f_\de$ to smooth $\hat{V}$: 
$$
\tilde{V}_{\si,\si'}(r) := \int f_\de(r-t)\hat  V_{\si,\si'}(t) dt. 
$$
Then $\tilde{V}_{\si,\si'}$ is twice continuously differentiable 
with uniformly bounded derivatives (since $\hat V$ is uniformly bounded and 
$f_\de$ has uniformly bounded derivatives).  For $s \ge r^{(0)}+ \ep_0$ we have 
$$
\tilde{V}_{\si,\si'}(r) -  V_{\si,\si'}(r) \ge 
\int \! f_\de(r-t) (\hat V_{\si,\si'}(t) - V_{\si,\si'}(r)) dt \ge 0, 
$$ 
since for $t \in [r-\de,r+\de]$ we either have 
$\hat V_{\si,\si'}(t) = M$ or $|V_{\si,\si'}(t)-  V_{\si,\si'}(r)| \le L|r-t| \le L\de$. 
Similarly it can be seen that 
$\tilde{V}_{\si,\si'}(r) -  V_{\si,\si'}(r) \le 2 \de L$
for $r \in (r^{(i-1)}_{\si,\si'} +\ep + \de,r^{(i)}_{\si,\si'} - \ep - \de)$ and arbitrary $i$. 
Denoting $F_\de$ the cumulative distribution function of $f_\de$ we set 
\begin{align*}
\bar{V}_{\si,\si'}(r) := \tilde{V}_{\si,\si'}(r) (1-F_\de(r -R-\de)) + 
V_{\si,\si'}(r) F_{\de}(r-R-\de). 
\end{align*}
for $r > r^{(0)}_{\si,\si'} + \ep_0$. We note that $\bar{V}_{\si,\si'}(r) = V_{\si,\si'}(r)$ for 
$r \ge R+ 2 \delta$ and  $\bar{V}_{\si,\si'}(r) = \tilde V_{\si,\si'}(r)$ for $r \le R$;  
in particular $|\bar V''_{\si,\si'}(r)| \le \frac{c}{r^{\al}}$ and 
and $|\bar V'_{\si,\si'}(r)| \le \frac{c}{(\al-1) r^{\al-1}}$ for $r > R+2 \de$, 
and for $r \le R+2\de$ the derivatives are uniformly bounded. 
Finally we let $\bU_{\si,\si'}(x) := \bar{V}_{\si,\si'}(\|x\|)$ for $x \in K^c$ and $u:= \bU - U$. 
Then \eqref{equ:smoothU} is satisfied: 
Note that $|\partial_{\ee}^2 \bU_{\si,\si'}(x)| \le 
|\bar V''_{\si,\si'}(\|x\|)| + \frac 2 {\|x\|} |\bar V'_{\si,\si'}(\|x\|)| \le \frac{c'}{\|x\|^{\al}}$
for some sufficiently large $c'$, 
and thus $|\partial_{\ee}^2 \bU_{\si,\si'}(x+t\ee)| \le \frac{c''}{\|x\|^{\al}} =:\psi(x)$ 
for all $t \le \frac{\|x\|} 2$ and $c'' := c' 2^{\al}$. 
Furthermore
$$
\int 1_{K^c}(x,\si,\si') \psi(x)(1 \vee \|x\|^2) dx 
\le \int_{\ep_0}^1 \frac{2\pi c''}{r^{\al -1}} dr + \int_{1}^\infty  \frac{2\pi c''}{r^{\al-3}} dr 
< \infty.  
$$
Also \eqref{equ:smoothu} is satisfied: 
By construction $u \ge 0$ on $K^c$ and $u_{\si,\si'}(x) = 0$ for $\|x\|> R +\ep$ and 
\begin{align*}
&\int \! 1_{K^c}(x,\si,\si') (u_{\si,\si'}(x) \wedge 1) dx 
\le N \pi 8R\ep + 2 \de L (R+\ep)^2\pi < \ga
\end{align*}
by choice of $\ep$ and $\de$, 
since $u \wedge 1 \le 1$ in the $N$ circular regions near $r^{(i)}_{\si,\si'}$,  
and $u \wedge 1 \le 2 \de L$ in the complement of these regions. 
\qed

\subsection{Ruelle bound: \autoref{Lem:Ruellebound}}

\autoref{Lem:Ruellebound} is identical to Lemma 6 of \cite{R2}, and its proof is 
included here only to make the presentation self contained. 
In the setting of \autoref{Lem:Ruellebound} let 
$\mu \in \G_{\YY'}^{U,\be,z}$, $m \ge 0$ and $f:(\bR^2_S)^m \to [0,\infty)$. 
For every $g:\YY \to [0,\infty)$ and $Y' \in \YY'$ we have 
\begin{align*}
\int & \nu_{\La_n}(dY|Y') \sumn_{y_1,\ldots,y_m\in Y_{\La_n}}f(y_1,\ldots,y_m)g(Y) \\
&=\int_{\La_n^m} dy_1\cdots dy_m f(y_1,\ldots,y_m)\int \nu_{\La_n}(dY|Y') 
g(\{y_1,\ldots,y_m\}\cup Y),
\end{align*}
which follows from the definition of the Poisson point process. 
Setting $g(Y)=\frac1{Z_{\La_n}^{U,\be,z}(Y')} e^{-\be H_{\La_n}^U(Y)}z^{|Y_{\La_n}|}$  
and integrating the equation w.r.t. $\mu(dY')$ we get 
\begin{align*}
&\int \mu(dY) \sumn_{y_1,\ldots,y_m\in Y_{\La_n}}f(y_1,\ldots,y_m) = \\
&= \int_{\La_n^m} \! dy_1\cdots dy_m f(y_1,\ldots,y_m)\int \! \mu(dY) 
e^{-\be (H^U(\{y_1,...,y_m\})+W^U(\{y_1,...,y_m\},Y))} z^m\\
&=\int_{\La_n^m} \! dy_1\cdots dy_m f(y_1,\ldots,y_m)  \rho_\mu^{U,\be,z}(\{y_1,...,y_m\})
\end{align*}
By definition of the Ruelle bound we have $\rho_\mu^{U,\be,z}(\{y_1,...,y_m\}) \le \xi^m$ and letting $n \to \infty$ the assertion follows via monotone convergence. 
\qed

\subsection{Properties of the shift function: \autoref{Lem:transproperties}}

Recall that we call a function $f: \R^2_S \to \R$ $\de$-Lipschitz continuous if 
$$
\forall x,x' \in \R^2,  \si \in S: |f(x,\si) - f(x',\si)| \le \de \|x-x'\|. 
$$

\begin{Lem}\label{Lem:Lip}
For all $y \in \R^2_S, \tau \ge 0$ $\; t_0 \wedge \fm_{y,\tau}$ is $\de$-Lipschitz continuous. 
\end{Lem}

\Pf We first note that for all $s > 0$ we have 
$
0 \ge \ftn'(s) \ge - \frac{3\de^2}{\sqrt{\log n}} \frac 1 {n^{2/3}}
\ge - \de.
$
$t_0 = \ftn(\|.\|_\infty)$ is thus $\de$-Lipschitz continuous. 
So in case  of $h_{y,\tau} > \de \ep$ we are done. 
In case of $h_{y,\tau} \le \de \ep$ we note that 
$\fm_{y,\tau}$ is $\de$-Lipschitz continuous wherever it is finite, 
since $d_K(y,.)$ is $1$-Lipschitz continuous by definition of $d_K$ and $K_\ep$. 
Thus it suffices to show that at the boundary points $y'$, 
where $\fm_{y,\tau}(y')$ goes from finite to infinite values, 
the function values of the distortion are above $\ftn(\|y'\|_\infty)$. 
For $(y,y'') \in K_\ep$ we have $d_K(y,y'') \le \ep$ and 
for $(y,y'') \in K_\ep^c$ we have $d_K(y,y'') \ge \ep$, 
which by continuity of $d_K(y,.)$ implies that $d_K(y,y') = \ep$ for such boundary points. 
So we get   
$$
\tau + \frac{h_{y,\tau}}\ep d_K(y,y') = \tau + |\ftn(\|y\|_\infty - c_K) - \tau| 
\ge \ftn(\|y\|_\infty - c_K) \ge \ftn(\|y'\|_\infty)
$$
as desired. Here we have used the definition of $h_{y,\tau}$, the monotonicity of $\ftn$
and the definition of $c_K$. \qed 

\bigskip 

We will now collect some properties of the construction of $\cTn$ 
that will be used here and later on. 
Let $t_k,P_k,C_k,\tau_k$ and $m$ as in the construction 
of $\cTn$. Furthermore let 
$$
T_k: \R^2_S \to \R^2_S, \; T_k(y) := y + t_k(y) \ee
$$
denote the $k$-step transformation function. On $\R^2_S$ 
we consider the partial order
$$
(x_1,x_2,\si) \le_{\ee} (x_1',x_2',\si') \quad :\Lra \quad 
x_1 \le x_1' ,x_2 = x_2', \si = \si'. 
$$

\begin{Lem}\label{Lem:constrproperties}
For the above objects we have for all $k \ge 0$: 
\begin{align} 
&\tau_k \le \tau_{k+1} \quad \text{ and } \quad  t_k \ge t_{k+1}, \label{equ:mono}\\
&t_{m+1}(y) = \tau_k \quad \text{ for } y \in C_k, \label{equ:transm}\\
&t_k \text{ is $\de$-Lipschitz continuous},\label{equ:tLip}\\
&T_k \text{ is bijective and $\le_{\ee}$-increasing}, \label{equ:Tcont}\\ 
&T_k^{-1}(y) + t_k(T_k^{-1}(y)) \ee = y \text{ for all } y \in \R^2_S \label{equ:inv}\\
&t_k \circ T_k^{-1} \ge t_{k+1} \circ T_{k+1}^{-1}.\label{equ:monocomp}
\end{align}
\end{Lem}
\Pf
For the first part of \eqref{equ:mono} we note that for $y \in P_{k+1}$
$$
\tau_{k+1} = t_{k+1}(y) = t_k(y) \wedge \bigwedge_{y' \in C_k} \fm_{y',\tau_k}(y)
\ge \tau_k,
$$
where the equalities are by definition of $\tau_{k+1}, t_{k+1}, t_k$ and 
the inequality is by choice of $P_k$ and the definition of $\fm_{y',\tau_k}$. 
The second part of \eqref{equ:mono} is clear from the definition of $t_k$.
For \eqref{equ:transm} we note that for $y \in C_k$ we have 
$$
t_{m+1}(y) = t_k(y) \wedge \bigwedge_{k \le j \le m} \bigwedge_{y' \in C_j} 
 \fm_{y',\tau_j}(y)
$$
and $t_k(y) \ge \tau_k$ by choice of $\tau_k$, $\fm_{y,\tau_k}(y) = \tau_k$ and 
$\fm_{y',\tau_j}(y) \ge \tau_j \ge \tau_k$ for all $j \ge k$ and $y' \in C_j$ by \eqref{equ:mono}. 
\eqref{equ:tLip} is an immediate consequence of \autoref{Lem:Lip} by definition 
of $t_k$ since the minimum of $\de$-Lipschitz continuous functions is again $\de$-Lipschitz continuous. 
For \eqref{equ:Tcont} we note that for every $x_2  \in \R, \si  \in S$,
the function 
$T_k: A_{x_2,\si} := \{(x_1,x_2,\si): x_1 \in \R\} \to A_{x_2,\si}$
is continuous and strictly $\le_{\ee}$-increasing by \eqref{equ:tLip} and thus bijective.  
\eqref{equ:inv} follows immediately from the definition of $T_k$. 
For \eqref{equ:monocomp} we note that $T_{k+1} \le_{\ee} T_k$ by \eqref{equ:mono}. 
This implies that $T_k^{-1} \le_{\ee} T_{k+1}^{-1}$ by 
\eqref{equ:Tcont}, and using \eqref{equ:inv} we get \eqref{equ:monocomp}.
\qed 

\bigskip

Now we prove \autoref{Lem:transproperties}.
For \eqref{equ:transLan} we note  that $\Tn(y) = T_{m+1}(y)$ for all $y \in Y$ 
by \eqref{equ:transm}, and $T_{m+1}(y) = y$ for $y \in \La_n^c$ by 
definition of $t_{m+1}$ and thus the bijectivity of $T_{m+1}$ from \eqref{equ:Tcont}
also gives $T_{m+1}(y) \in \La_n$ for $y \in \La_n$. 
For \eqref{equ:transconstant} the case of $yy' \in B$ is clear,  
since $B$-clusters are shifted by the same amount. If $(y,y') \in K$
we may assume that $y \in C_k$, $y' \in C_l$, w.l.o.g. $k < l$.
Since $t_l$ contains the distortion $\fm_{y,\tau_k}$ and 
$\fm_{y,\tau_k}(y') = \tau_k$ by definition of $\fm_{y,\tau_k}$, 
we have $\tau_l \le t_l(y') \le \tau_k$ and thus $\tau_l = \tau_k$.
For the proof of \eqref{equ:transLip} we may assume that $y \in C_k, y' \in C_l$, 
w.l.o.g. $k < l$. Let $y''$ be the particle with the position of $y$ and the spin of $y'$. 
We have 
$$
0 \le \tn(y') - \tn(y) \le t_{m+1}(y') - t_{m+1}(y'') \le \de \|y'-y''\| = \de \|y'-y\|,
$$
where we have used $\tau_k \le \tau_l$ by \eqref{equ:mono}, 
$\tau_l = t_{m+1}(y')$ by \eqref{equ:transm}, 
$t_{m+1}(y'') \le \tau_k$ (which is due to $\fm_{y,\tau_k}(y'') = \tau_k$), 
\eqref{equ:tLip} and the choice of $y''$. 
\qed

\subsection{Properties of good configurations: \autoref{Lem:goodprop}}

We have the following estimates of the translation distance: 

\begin{Lem}\label{Lem:translowerbound}

Let $\bY = (Y,B) \in \bYY$. For every $y \in Y$ we have $\tn(y) \le \ftn(\|y\|_\infty)$,  
and 
$$
\forall k \le m^*, y \in C_k \exists y' \in C_{B_+}^\downarrow(y): \tn(y) \ge \ftn(\|y'\|_\infty), 
\quad \text{ where }
$$
$C_{B_+}^\downarrow(y) := \{y' \in Y \!: \!\exists k_0 \ge ... \ge k_l , y_i \in C_{k_i} \!: 
y = y_0, y' = y_l, \forall i : y_{i-1}y_i  \in  B_+\}$.
Furthermore $y'$ can be chosen such that $\|y'\|_\infty \ge \|y\|_\infty$. 
\end{Lem}

\Pf For the first estimate observe that for $y \in C_k$ we have 
$\tau_k \le t_k(y) \le t_0(y)$ by definition of $P_k$ and \eqref{equ:mono}. 
For the second estimate let $y \in  C_k$ with $k \le m^*$ and consider a sequence $y=y_0,...,y_l$ 
such that $y_i \in C_{k_i}$, $k_0 \ge ... \ge k_l$, $y_{i-1}y_i  \in  B_+$
with minimal $k_l$. Since $C_{k_l} = C_B(P_{k_l})$ we may assume that $y_l \in P_{k_l}$. 
By \eqref{equ:mono} we have $\tn(y_0) \ge \tn(y_l)$ and by choice of $k_l,y_l$ and using 
$k_l \le k_0 \le m^*$ we have  $\fm_{C_{k'},\tau_{k'}}(y_l) = \infty$ for all $k' < k_l$ 
and thus $\tn(y_l) = t_0(y_l)$. Thus $y' := y_l$ is a suitable choice.  
For the last statement consider the case $\|y'\|_\infty \le \|y\|_\infty$, 
then $\ftn(\|y'\|_\infty) \ge \ftn(\|y\|_\infty)$, thus we may replace $y'$ by $y$.\qed  

\bigskip

For the proof of \autoref{Lem:goodprop} let $\bY = (Y,B) \in \Gn$. 
For the first assertion let $y \in C_k$ for $k := m^*$. 
By \autoref{Lem:translowerbound} 
there is a $y' \in C_{B_+}^\downarrow(y) \subset C_{B_+}(y)$ such that 
\begin{equation}\label{equ:esty}
\ftn(\|y\|_\infty)   \ge \tn(y) \ge \ftn(\|y'\|_\infty) \ge 
\ftn(\frac{1 \! \vee  \!\|y\|_\infty \log \|y\|_\infty}{\de}),   
\end{equation}
where we have also used the monotonicity of $\ftn$ and $\bY \in \Gn$.  
Thus 
$$
h_{y,\tau_k} \le \ftn(s \!- \! c_K) - \ftn(\frac{1 \vee s \log s}{\de})
=: F(s) \quad \text{ for } s := \|y\|_\infty.
$$
We claim that $F(s) \le \de \ep$, 
which implies $m^* = m$ by definition of $m^*$. 
For this claim we note that $F(s) \le F(n^{2/3})$ for $s\in[0,n^{2/3}]$
and $F(s) \le F(n)$ for $s \in [n,\infty)$ (by monotonicity of $\ftn$). 
For $s \in [n^{2/3}, n]$ we have 
$$
F(s) 
\le  \frac{3c }{\sqrt{\log n}} \log \frac{s \log s}{\de(s-c_K)}
\le \frac{3 \de^2}{\sqrt{\log n}} \log \frac{\log n}{\de(1- \frac{c_K}{n^{2/3}})}
\le \frac{3 \de^2}{\sqrt{\log n}} 2 \sqrt{\frac{2\log n}{\de}}
.  
$$
In the last step we have used $\de \le \frac 1 {c_K}$, which implies $\de(1-\frac{c_K}{n^{2/3}}) 
\ge \de - \frac 1 {n^{2/3}} \ge \de - \de^{16/3} \ge \frac \de 2$, 
and $\log x \le 2 \sqrt x$ for $x \ge 1$. We conclude that 
$F(s) \le \de \ep$ provided that $\de$ is sufficiently small. 

For the second assertion let $y \in Y$ such that $Y \in \La_{\sqrt n}$.
By choice of $n \ge N = (\frac 1 \de)^8$ and $\de \le 10^{-6}$ we have 
$$
\frac{1 \vee \|y\|_\infty \log \|y\|_\infty}{\de} 
\le \frac {\sqrt n \log \sqrt n} \de  \le n^{2/3}, 
$$
so using \eqref{equ:esty} and the definition of $\ftn$
we have $\tn(y) = \ftn(\|y\|_\infty) = c \sqrt{\log n}$
as desired.  \qed

\subsection{Bijectivity of the transformation: \autoref{Lem:bijective}}

We note that \autoref{Lem:bijective} can be proved by combining the arguments 
of the proof of Lemma 12 in \cite{R2} and Lemma 3 in \cite{R3}, 
but we include the proof here for the sake of completeness. 
We will only show that $\cTn$ is bijective - $\cTn^-$ can be treated similarly. 
For defining an inverse transformation for $\cTn$ 
our main task is to reconstruct the partition 
of a configuration when given only the transformed image of the configuration. 
The following lemma solves this reconstruction problem: 
\begin{Lem} \label{lem:reconstruct} Let $\bY = (Y,B) \in \bYY$, $(Y',B') := \bY' := \cTn(\bY)$, 
$\tP_k:= P_k + \tau_k \ee$ and $\tC_k := C_k + \tau_k \ee$ for $0 \le k \le m$. 
$\tC_0$ is the $B'$-Cluster of $Y'_{\La_n^c}$. 
For every $k \ge 1$ $\tP_k$ is the set of points of $Y' \stm \bigcup_{j < k} \tC_j$, 
at which the minimum of $t_k \circ T_k^{-1}$ is attained,  
and $\tC_k$ ist the $B'$-Cluster of $\tP_k$. 
\end{Lem} 
\Pf The assertion for $\tC_0$ follows from $\tC_0 = C_0$ and $Y'_{\La_n^c} = Y_{\La_n^c}$, 
which is due to \eqref{equ:transLan}.
Now let $1 \le k \le l \le m$, $y_k' \in \tP_k$ and $y_l' \in \tC_l$. 
Then $y_k := y_k' - \tau_k \ee \in P_k$ and $y_l := y_l' - \tau_l \ee \in C_l$ 
and by definition and \eqref{equ:transm} we get $t_k(y_k) = \tau_k$, 
$T_k(y_k) = y_k'$, $t_{m+1}(y_l) = \tau_l$ and 
$T_{m+1}(y_l) = y_l'$.   
Using  \eqref{equ:mono} and \eqref{equ:monocomp} we deduce
$$
t_k(T_k^{-1}(y_k')) = \tau_k \le \tau_l = t_{m+1}(T_{m+1}^{-1}(y_l'))
\le t_k(T_k^{-1}(y_l')).
$$
In case of $t_k(T_k^{-1}(y_k')) = t_k(T_k^{-1}(y_l'))$ we must have  
$\tau_k = \tau_l$ and $\tau_l = t_k(T_k^{-1}(y_l'))$, 
which implies $T_k^{-1}(y_l') = y_l' - \tau_l \ee = y_l$ by \eqref{equ:inv}, i.e. 
$T_k(y_l) = y_l'$ and thus $t_k(y_l) = \tau_l = \tau_k$. By definition of $P_k$ this 
implies $y_l \in P_k$, i.e. $y_l' \in \tP_k$. \qed 

\bigskip 

This motivates the following definition of the inverse transformation.  
For $\bY'= (Y',B') \in \bYY$ we recursively define 
$\tt_k,\tT_k,\tP_k,\ttau_k,\tC_k$ (depending on $n,c,\bY'$): 
For $k = 0$ we let $\tt_0 := \ftn(\|.\|_\infty)$, $\tC_0$ be the $B'$-cluster of 
$Y'_{\La_n^c}$ and $\ttau_0 := 0$. For the recursive step let $k \ge 1$ and 
define $\tt_k: \R^2_S \to [0,\infty)$ and $\tT_k: \R^2_S \to \R^2_S$ by
$$
\tt_k := \tt_0 \wedge \bigwedge_{l<k} \bigwedge_{y \in \tC_l} \fm_{y - \ttau_l \ee, \ttau_l} 
\quad \text{ and } \quad \tT_k(y) := y + \tt_k(y) \ee, 
$$ 
let  $\tP_k$ be the set of points of $Y' \stm \bigcup_{l < k} \tC_l$ 
at which the minimum of $\tt_k \circ \tT_k^{-1}$ is attained and   
$\ttau_k$ be the corresponding minimal value. 
(We will show that this is well defined in that $\tT_k$ is invertible.) 
Let $\tC_k$ the $B'$-Cluster of $\tP_k$. 
This finishes the recursive step. Let $\tm$ be the smallest value such that 
$\bigcup_{l=0}^{\tm} \tC_l = Y'$, and let 
\begin{align*}
&\text{
$\ttn(y) = \ttau_k$ and $\tTn(y) =  y - \ttau_k \ee$ for all $y \in \tC_k$ and $1 \le k \le \tm$, 
}\\
&\text{ and } \tcTn(\bY') \, := \, (\{\tTn(y):y \in Y'\}, \{\tTn(y)\tTn(y'): yy' \in B'\}). 
\end{align*}
\begin{Lem} \label{lem:invwelldef}
Let $\bY' = (Y',B') \in \bYY$. For all $k \ge 0$ we have $\ttau_k \ge 0$ and 
\begin{align}
&\tT_k \text{ is bijective and $\le_{\ee}$-increasing}. \label{equ:Tcontinv}
\end{align} 
In particular the above construction is well defined. 
\end{Lem}
\Pf 
We proceed by induction on $k$. The case $k = 0$ follows from $\tt_0 = t_0$ and $\ttau_0 = 0$. 
For $k > 0$ the $\de$-Lipschitz continuity of $\tt_k$ 
follows from \autoref{Lem:Lip} by definition of $\tt_k$ 
(since $\ttau_l \ge 0$ for all $l < k$). 
From this we immediately get the desired properties of $\tT_k$, 
and $\ttau_k \ge 0$ follows from the definition of $\tt_k$ and $\ttau_l \ge 0$ for all $l < k$. 
\qed

\begin{Lem} \label{lem:eiginv}
Let $\bY' = (Y',B') \in \bYY$. For all $k \ge 0$
\begin{align}
&\tT_k^{-1}(y) + \tt_k(\tT_k^{-1}(y)) \ee   = y \quad \text{ for every } y \in \R^2_S, 
\label{equ:invinv} \\
&\ttau_k \le \ttau_{k+1}  \quad \text{ and } \quad 
\tt_k \ge \tt_{k+1}, \label{equ:moninv}\\
&\tT_{\tm+1}^{-1}(y) = y - \ttau_k \ee \quad \text{ for } y \in \tC_k. \label{equ:transminv}
\end{align} 
\end{Lem}

\Pf \eqref{equ:invinv} immediately follows from the definition of $\tT_k$.  
Using \eqref{equ:invinv} and the $\le_{\ee}$-monotonicity from \eqref{equ:Tcontinv} we obtain for all  $y \in \R^2_S$, $c \in \R$: 
\begin{equation}
\begin{split}
&\tt_k(\tT_k^{-1}(y)) \ge c  \quad \Leftrightarrow \quad  
y \ge_{\ee} \! \tT_k^{-1}(y) + c\ee \\
&\Leftrightarrow 
\tT_k(y - ce_1) \ge_{\ee} \! y  \quad \Leftrightarrow \quad 
\tt_k(y - c\ee) \ge c. \label{equ:Tcinv}
\end{split}
\end{equation}
The second part of \eqref{equ:moninv} follows immediately from the 
definition of $\tt_k$. 
For the first part of \eqref{equ:moninv} let $y \in \tP_{k+1}$. 
By definition of $\tP_k$ we have  $\tt_k(\tT_k^{-1}(y)) \ge \ttau_k$ 
and thus $\tt_k(y - \ttau_k \ee) \ge \ttau_k$ by \eqref{equ:Tcinv}. 
By definition of $\tt_{k+1}$ this implies $\tt_{k+1}(y - \ttau_k \ee) \ge \ttau_k$
and thus $\ttau_k \le \tt_{k+1}(T_{k+1}^{-1}(y)) = \ttau_{k+1}$ by \eqref{equ:Tcinv}
and definition of $\ttau_{k+1}$. 
For \eqref{equ:transminv} let $y \in \tC_k$.  By definition of $\tt_{m+1}$ we have 
$$
\tt_{m+1}(y-\ttau_k \ee) = \tt_k(y-\ttau_k\ee) \wedge \bigwedge_{k \le l \le m} \bigwedge_{y' \in \tC_l} \fm_{y' - \ttau_l \ee, \ttau_l}(y -\ttau_k\ee) = \ttau_k, 
$$
since $\tt_k(y-\ttau_k\ee) \ge \ttau_k$ (by \eqref{equ:Tcinv} and definition of $\tP_k$), $\ttau_l \ge \ttau_k$ for $l \ge k$ 
by \eqref{equ:moninv} and $\fm_{y-\ttau_k\ee,\ttau_k}(y-\ttau_k\ee) = \ttau_k$. 
This implies $\tT_{m+1}(y - \ttau_k \ee) = y$. \qed 

\bigskip 

We now can show  the analogue of the reconstruction result from \autoref{lem:reconstruct}.
\begin{Lem} \label{lem:reconstructinv}
Let $\bY' = (Y',B') \in \bYY$, $\tt_k, \tT_k, \tP_k, \tC_k$ and  $\ttau_k$ as above. 
Let $\bY := (Y,B) := \tcTn(\bY')$, $P_k := \tP_k - \ttau_k \ee$ and $C_k := \tC_k - \ttau_k \ee$ for $0 \le k \le \tm$. 
$C_0$ is the $B$-Cluster of $Y_{\La_n^c}$. 
For every $k \ge 1$ $P_k$ is the set of points of $Y \stm \bigcup_{j < k} C_j$ at which the minimum of $\tt_k$ is attained, and $C_k$ is the $B$-cluster of $P_k$. 
\end{Lem} 
\Pf We have that $\tT_{m+1}(y) = y$ for $y \in \La_n^c$ and by \eqref{equ:Tcontinv} this 
implies that $\tT_{m+1}^{-1}(y) \in \La_n$ for $y \in \La_n$. Using 
\eqref{equ:transminv} this implies $Y_{\La_n^c} = Y'_{\La_n^c}$. 
Thus $C_0 = \tC_0$ is the $B$-cluster of $Y_{\La_n^c}$. 
For $1 \le k \le l \le \tm$ let $y_k \in P_k$ and $y_l \in C_l$. 
Then $y'_k := y_k + \ttau_k \ee \in \tP_k$ and $y'_l := y_l + \ttau_l \ee \in \tC_l$
and by definition of $\tP_k$, \eqref{equ:invinv} and \eqref{equ:transminv} we get 
$\tt_k(\tT_k^{-1}(y'_k)) = \ttau_k$, $\tT_k^{-1}(y'_k) = y'_k - \ttau_k \ee = y_k$, 
$\tT_{\tm+1}^{-1}(y'_l) = y'_l - \ttau_l \ee = y_l$ and $\tt_{\tm +1}(\tT_{\tm+1}^{-1}(y'_l)) = \ttau_l$.  
Thus 
$$
\tt_k(y_k) = \ttau_k \le \ttau_l = \tt_{\tm+1}(y_l)  \le \tt_k(y_l)
$$
using \eqref{equ:moninv}. In case of $\tt_k(y_k) = \tt_k(y_l)$ we must have 
$\ttau_k = \ttau_l$ and $\tt_k(y_l) = \ttau_l$, which implies 
$\tT_k(y_l) = y'_l$ and thus $\tt_k(\tT_k^{-1}(y_l')) = \ttau_l = \ttau_k$, 
which gives $y_l' \in \tP_k$ by definition of $\tP_k$ and thus $y_l \in P_k$. \qed  

\bigskip 

\begin{Lem}\label{lem:inverse}
On $\bYY$ we have $\tcTn \circ \cTn  =  id$ and  $\cTn \circ \tcTn  =  id$.
\end{Lem}
\Pf 
For the first part let $\bY := (Y,B) \in \bYY$ and $\bY' := (Y',B') := \cTn(\bY)$.
Let $t_k,T_k,P_k,C_k,\tau_k$ be the corresponding objects in the construction of 
$\cTn(\bY)$ and $\tt_k,\tT_k,\tP_k,\tC_k, \ttau_k$ the corresponding objects
in the construction of $\tcTn(\bY')$. 
By \autoref{lem:reconstruct} we have $\tC_0 = C_0$ and by \autoref{lem:reconstruct}
and the construction of $\bY' = \cTn(\bY)$ inductively we obtain 
\begin{equation}  \label{tildegleichhc}
\tt_k = t_k, \; \tT_k = T_k, \; \ttau_k = \tau_k, \; \tP_k = P_k +\tau_k\ee \text{ and }
\tC_k = C_k + \tau_k \ee 
\end{equation} 
for every $k \ge 1$. By definition of $\tcTn$ we obtain $\tcTn(\bY') = \bY$. 
The second part follows similarly from \autoref{lem:reconstructinv}. \qed

\subsection{Density of the transformation:\autoref{Lem:density}}

\autoref{Lem:density} can be proved by combining the arguments used in the proofs of Lemma 13 from \cite{R2} and Lemma 4 from \cite{R3}, 
but we include the proof here for the sake of completeness. 
We present a simplified and shortened version of the proof.  
Again we will concentrate on $\cTn$. 
The corresponding property of $\cTn^-$ can be shown similarly. 

\begin{Lem} \label{Lem:Bernoulli}
For every $\mu \in \G$ we have $H_\La^u < \infty$ $\mu$-a.s. and 
for every measurable $f: \bYY \to [0,\infty)$ 
\begin{align*}
\int \mu \otimes \pi_n(d\bY) f(\bY) = \int \mu(dY) e^{-\be H^u_{\La_n}(Y)} 
\sum_{B \Subset E_{\Lan}(Y)}  \prod_{b \in B} (e^{\be u(b)}-1) f(Y,B),
\end{align*}
where $B \Subset E$ is a shorthand notation for $B \subset E: |B| < \infty$. 
\end{Lem}

\Pf
 Let $u_\wedge := \be u \wedge 1$. Since 
$\sum \limits_{b \in E_{\La_n}(Y)} \! u_\wedge(b) 
\le  \sumn \limits_{y,y' \in Y} 1_{\{y \in \La_n\}} u_\wedge(y,y')$
\begin{align*}
\int \mu(dY) \!&\sum_{b \in E_{\La_n}(Y)} \! u_\wedge(b) 
\le \xi^2 \int_{\La_n}  \!\! dy \int dy' u_\wedge(y,y')  
\le \xi \int_{\La_n} \!\! dy  = \xi 4n^2  
< \infty, 
\end{align*}
where we have used \autoref{Lem:Ruellebound} in the first step 
and \eqref{equ:constants} in the second. 
Thus $\sum \limits_{b \in E_{\La_n}(Y)} \!\! u_\wedge(b)< \infty$ a.s., which implies
$\be H^u_{\La_n}(Y) = \!\! \sum \limits_{b \in E_{\La_n}(Y)} \!\!\! \be u(b) < \infty$ a.s. 
(since $u_\wedge(b) < 1$ for all but finitely many edges $b$). 
This gives the first assertion. Since 
$$
\sum_{b \in E_{\La_n}(Y)} 1-e^{-\be u(b)}
\le \sum_{yy'\in E_{\La_n}(Y)} \be u(y,y') = \be H^u_{\La_n}(Y) < \infty, 
$$
by Borel-Cantelli for $\mu$-a.e. $Y$ $\pi_n(.|Y)$ is concentrated on finite edge sets.
Since there are only countably many of those we are done by 
\eqref{equ:Bernoulli}. \qed

\bigskip 
 
For the proof of \autoref{Lem:density} let $\mu \in \G$ and $f: \bYY \to [0,\infty)$ 
be measurable.
In order to be able to fix edge configurations before specifying particle positions  
we define $Y_k := \{1,...,k\} \cup Y_{\La_n^c}$ and $E_n(Y_k) := \{yy': y \in\{1,...,k\}, y' \in Y_k\}$. 
For $B \subset E_n(Y_k)$ and $y \in \La_n^k$ we define $\bY_y := (Y_y,B_y) \in \bYY$ to be the particle-edge configuration, 
where every $i \in \{1,...,k\}$ is replaced by the particle $y_i$. 
To simplify notation we will write $\bY_y$ both for $Y_y$ and $B_y$. 
Using the above lemma, $\mu = \mu \otimes \mu_n$, the definition of $\mu_n$ and the 
Poisson point process $\nu_n(.|Y)$ we get 
\begin{align*}
\int \mu \otimes \pi_n&(d\bY) f(\bY) 
= 
\int \mu(dY) \sum_{k \ge 0} \sum_{B \Subset E_n(Y_k)} \frac {e^{-4n^2}z^k} {Z_n(Y) k!}   
\int_{\La_n^k} dy  \tilde{f}(\bY_y), \\
&\text{ where } \tilde{f}(\bY_y) :=  \prod_{b \in \bY_y} (e^{\be u(b)}-1)  e^{-\be H^{\bU}_{\La_n}(\bY_y)} f(\bY_y).  
\end{align*} 
Thus it suffices to show 
for every $Y \in \YY$, $k \ge 0$ and $B \Subset E_n(Y_k)$ we have 
\begin{align*}
&\int_{\La_n^k} dy  \prod_{b \in \bY_y} (e^{\be u(b)}-1) e^{-\be H^{\bU}_{\La_n}(\bY_y)}  \ph_n(\bY_y) f(\cTn(\bY_y))  = 
\int_{\La_n^k} dy \tilde{f}(\bY_y).
\end{align*}
$u$ is translation invariant and by \eqref{equ:transconstant} particles connected 
by an edge are shifted by the same distance, so 
$\prod_{b \in \bY_y} (e^{\be u(b)}-1) = \prod_{b \in \cTn(\bY_y)} (e^{\be u(b)}-1)$, 
thus by definition of $\ph_n$ it suffices to show that 
$$
\int_{\La_n^k} dy \th_n(\bY_y) \tilde{f}(\cTn(\bY_y)) = \int_{\La_n^k} dy  \tilde{f}(\bY_y).
$$
From now on $Y$, $k$ and $B$ are fixed and will be suppressed in notations. 
We would like to view $\cTn(\bY_y)$ as a transformation of $y$. 
So let us define 
$$
T(y) := y' \quad \text{ such that } y_i' = y_i + \tau_j \ee \text{ for } y_i \in C_j.
$$
(Here and later we consider $t_j,T_j,\tau_j,P_j,C_j,m$ from the construction of $\cTn(\bY_y)$.)
We note that by \eqref{equ:transLan} we have $T: \La_n^k \to \La_n^k$, by the last subsection 
$T$ is bijective,  and by definition $\cTn(\bY_y) = \bY_{T(y)}$. 
It will be useful to distinguish the order, in which 
the points of $\bY_y$ are shifted. Let $\Pi$ denote the set of all finite sequences 
$\eta = (\eta_1,...,\eta_m)$ of disjoint nonempty subsets of $\{1,...,k\}$ such 
that the corresponding $B$-clusters $\eta_j^B$ ($1 \le j \le m$) together 
with $\eta_0^B$, the $B$-cluster of $Y_{\La_n^c}$, give a partition of $Y_k$. 
We set $\eta'_j := \eta_j^B \stm \eta_j$. 
Let $m(\eta) := m$ be the length of the sequence $\eta$. For $\eta \in \Pi$ let 
\begin{align*}
A_\eta &:= \{y \in \La_n^k: m(\eta) = m, \forall j \ge 1: P_j = \{y_i: i \in \eta_j\}\} \text{ and }\\
{\tilde A}_\eta &:= \{y' \in \La_n^k: m(\eta) = \tm, \forall j \ge 1: \tP_j = \{y_i': i \in \eta_j\}\}.
\end{align*}
(Here and later $\tm,\tP_j,\tt_j, \ttau_j$ are the objects from the construction of $\tcTn(\bY_{y'})$.)  
%
Since both $A_\eta, \eta \in \Pi,$ and ${\tilde A}_\eta,\eta \in \Pi,$ 
give a disjoint decomposition of $\La_n^k$, it suffices to show that 
$$
\int_{\La_n^k} dy  \th_n(\bY_y) \tilde{f}(\bY_{T(y)}) 1_{A_\eta}(y) = \int_{\La_n^k} dy'   \tilde{f}(\bY_{y'}) 1_{{\tilde A}_\eta}(y')
$$
for every $\eta \in \Pi$. Let $\eta \in \Pi$ be fixed for the rest of the proof. 
We note that for all $y,y' \in \La_n^k$ such that $y' := T(y)$ and for all $j \le m$  
\begin{equation}\label{equ:Aj}
\forall  l \le j: P_l = \{y_i: i \in \eta_l\} \quad \Lra \quad  
\forall l \le j: \tP_l = \{y_i': i \in \eta_l\}.
\end{equation}
For '$\Rightarrow$' note that $P_1,...P_j$ determine $C_1,...,C_j$ (since $B$ is fixed) 
and recursively also $t_1,\tau_1,...,\tau_j$ and we are done by
the definition of $T$ and $\tP_l$. 
For '$\Leftarrow$' note that similarly $\tP_1,...,\tP_j$ determine 
$\tC_1,..,\tC_j$ and recursively also $\tt_1,\ttau_1,...,\ttau_j$. 
By the proof of \autoref{lem:inverse} we get $\ttau_1 = \tau_1,....\ttau_j = \ttau_j$ 
and thus we are done by the definition of $T$ and $\tP_l$. 
In particular \eqref{equ:Aj} implies that 
$$
y \in A_\eta \quad \Lra \quad T(y) \in \tilde{A}_\eta.
$$ 
Thus setting $g(y) := \tilde{f}(\bY_y) 1_{{\tilde A}_\eta}(y)$, 
using the above equivalence and the definition of $\th_n$, 
it suffices to show that 
$$
\int_{\La_n^k} dy  \prod_{j=1}^m \prod_{i \in \eta_j} |1+\partial_{\ee} t_j(y_i)| g(T(y))= \int_{\La_n^k} dy' g(y'). 
$$
Using Tonelli's theorem we now fix the order of integration on the left hand side (corresponding to the order given 
by $\eta$) to get 
$$
(\prod_{i \in \eta_0'} \int_{\La_n} dy_i )\Big(\prod_{j=1}^m ( \prod_{i \in \eta_j} \int_{\La_n} dy_i |1+\partial_{\ee} t_j(y_i)|)
(\prod_{i \in \eta_j'} \int_{\La_n} dy_i)\Big) g(T(y))
$$
and transform the integrals in the given order from $\eta_m'$ to $\eta_0'$. 
Whenever we consider $i \in \eta_j'$ we note that $y_i' = y_i + \tau_j \ee$ is 
a shift by a constant, because by \eqref{equ:Aj} we have 
$P_l = \{y_p: p \in \eta_l\}$ for all $l \le j$ and thus $\tau_j$ only depends on $y_i$ for $i \in \eta_0'\cup \eta_1 \cup ... \cup \eta'_{j-1} \cup \eta_j$, 
and thus $dy_i' = dy_i$. 
Whenever we consider $i \in \eta_j$ we note that $y_i' = T(y_i) = y_i + t_j(y_i)\ee$, 
and by \eqref{equ:Aj} we have $P_l = \{y_i: i \in \eta_l\}$
for all $l < j$ and thus $t_j$ only depends on $y_i$ for 
$i \in \eta_0'\cup \eta_1 \cup ... \cup \eta'_{j-1}$, 
and thus the Lebesgue transformation theorem 
gives $dy_i' = |1+\partial_{\ee} t_j(y_i)| dy_i$. 
Here we use that only the first coordinate of the particle is changed 
in the transformation (whereas the second coordinate and the spin remain the same)
and by the Lipschitz continuity from \eqref{equ:tLip} and Rademacher's theorem 
$\partial_{\ee} t_j$ exists a.e.. 
This establishes the above equation, which completes the proof 
of Lemma \autoref{Lem:density}. \qed

\subsection{Probability of good configurations: \autoref{Lem:goodprob}}

For $\bY = (Y,B) \in \Gn^c$ by definition we have $m \ge 1$ and distinct
$y_0,...,y_m \in Y$ such that $y_{i-1}y_i \in B_+$ and 
$$
\|y_m\|_\infty > \frac{1 \vee \|y_0\|_\infty \log  \|y_0\|_\infty} {\de} > \|y_0\|_\infty + 
\frac{1 \vee \|y_0\|_\infty \log  \|y_0\|_\infty} {2\de},
$$
since $\de < \frac 1 6$. 
By the triangle inequality 
$$
\|y_m\|_\infty - \|y_0\|_\infty \le 
\|y_m-y_0\|_\infty \le  m \|y_{k} - y_{k-1}\|_\infty
$$
for some $1 \le k \le m$. Combining the above inequalities we get 
\begin{align*}
1_{\Gn^c}(Y,B)  \le 
\sum_{m \ge 1} \sum_{k=1}^m \;\; \sumn_{y_0,...,y_m \in Y} \! \frac{(2 \de m)^2 \|y_{k} - y_{k-1}\|_\infty^2}{1 \vee \|y_0\|_\infty^2 (\log  \|y_0\|_\infty)^2} 
\prod_{i=1}^m 1_{\{y_{i-1}y_i \in B_+\}}.  
\end{align*}
We let $g := 1_{K_\ep \stm K^U} + \be u \wedge 1$ and note that for $\mu$-almost every $Y \in \YY$
\begin{equation} \label{equ:piint}
\int \pi_n(d(Y,B)|Y)\prod_{i=1}^m 1_{\{y_{i-1}y_i \in B_+\}}
\le \prod_{i=1}^m g(y_{i-1},y_i),   
\end{equation}
which follows from the definition of the Bernoulli measure and the hard core
and from $1-e^{-\be u(b)} \le \be u(b) \wedge 1$. 
Using this and \autoref{Lem:Ruellebound} we obtain that 
$$
\mu \otimes \pi_n(\Gn^c) \le \sum_{m \ge 1} \sum_{k=1}^m \int \! dy_0 \;... \! \int \! dy_m 
\frac{\xi (2 \de m)^2 \|y_{k} - y_{k-1}\|_\infty^2}{1 \vee \|y_0\|_\infty^2 (\log  \|y_0\|_\infty)^2} \prod_{i=1}^m \xi g(y_{i-1},y_i)
$$
By \eqref{equ:constants} we have 
\begin{equation} 
\label{equ:intg}
\xi \int g(y,y') dy' \le c_{u} \quad \text{ and } \quad  
\xi \int g(y,y') \|y'-y\|^2 dy' \le c_u',
\end{equation}
and we have 
$$
\int dy_0 \frac{1}{1 \vee \|y_0\|_\infty^2 (\log  \|y_0\|_\infty)^2}
=  \int_0^\infty \frac {8s} {1 \vee s^2(\log s)^2} ds \le 30.  
$$
Thus estimating the integrals above (starting with $\int dy_m$) we get 
$$
\mu \otimes \pi_n(\Gn^c) \le \sum_{m \ge 1} \sum_{k=1}^m  30 \xi (2 \de m)^2 c_u' c_{u}^{m-1} \le \de 
$$
if $\de$ is sufficiently small since $c_u < 1$ and thus 
$\sum_{m \ge 1} m^3c_{u}^{m-1} < \infty$. \qed

\subsection{Energy estimate: \autoref{Lem:energyestimate}}

We first note that by the triangle inequality and the symmetry of $\bU$ we have 
\begin{align*}
&S_1(\bY) \le \sum_{yy' \in E_{\La_n}(Y)} 
|\ph_{y,y'}(\vth_{y,y'}) + \ph_{y,y'}(-\vth_{y,y'}) - 2 \ph_{y,y'}(0)| \text{ where  } \\
&\ph_{y,y'}(t) := \bU(y,y' + t \ee) \quad \text{ and } \quad  
\vth_{y,y'} = \tn(y') - \tn(y).
\end{align*}
By Taylor expansion of $\ph_{y,y'}$ we get 
\begin{align*}
\ph_{y,y'}(\vth_{y,y'}) = \ph_{y,y'}(0) + \ph_{y,y'}'(0) \vth_{y,y'} + 
\frac 1 2 \ph_{y,y'}''(s \vth_{y,y'}) \vth_{y,y'}^2 
\end{align*}
for some $s \in [0,1]$ and likewise for $\ph_{y,y'}(-\vth_{y,y'})$. 
In the above estimate of $S_1$ the Taylor terms of order 0 and 1 cancel, 
thus 
$$
S_1(\bY) \le \sum_{yy' \in E_{\La_n}(Y)} 
\sup_{s \in [-1,1]} |\ph_{y,y'}''(s \vth_{y,y'})| \vth_{y,y'}^2
\le \sum_{yy' \in E_{\La_n}(Y)} \psi(y,y') \vth_{y,y'}^2. 
$$
For the last step we note that in case of $(y,y') \in K$ we have $\vth_{y,y'} = 0$ by \eqref{equ:transconstant}, 
and in case of $(y,y') \in K^c$ we may use the $\psi$-domination of $\bU$, since 
$|\vth_{y,y'}| \le \frac 1 2 \|y-y'\|$ by \eqref{equ:transLip} and 
$(y,y' +  s \vth_{y,y'}\ee) \in K^c$ for $s \in [-1,1]$,  
which will be shown at the end of this subsection. To estimate $\vth_{y,y'}^2$, 
w.l.o.g. $y \in C_l, y' \in C_k$ with $l > k$. If $k > m^*$ then 
$\tau_l = \tau_k = \tau_{m^*}$ by definition of $m^*$ and $t_{m^*+1}$ and 
thus $\vth_{y,y'} = 0$. If $k \le m^*$ we use \autoref{Lem:translowerbound} to obtain 
\begin{align*}
|\vth_{y,y'}| &= \tn(y) - \tn(y')
 \le \ftn(\|y\|_\infty) -  \ftn(\|y''\|_\infty)\\
&\le (\ftn(\|y\|_\infty) -  \ftn(\|y'\|_\infty)) +  
 (\ftn(\|y'\|_\infty)-  \ftn(\|y''\|_\infty)) 
\end{align*}
for some $y'' \in C_{B_+}^\downarrow(y')$ such that $\|y''\|_\infty \ge \|y'\|_\infty$, and
we have $m \ge 0$ and distinct $y_0,...,y_m \in Y$ (also distinct from $y$) 
s.t. $y_0 = y', y_m = y'', y_{i-1}y_i \in B_+$. 
Using $(a+b)^2 \le 2a^2+2b^2$ 
we can put everything together to estimate 
\begin{align*}
&S_1 \le S_1' + S_1'', \qquad \text{ where } \quad S_1' := \sumn_{y,y' \in Y} 2 \psi(y,y') \De(y,y'),\\
&S_1'' := \sum_{m \ge 1} \quad \sumn_{y,y_0,...,y_m \in Y} 2 \psi(y,y_0) 1_{\La_n}(y) \De(y_0,y_m) \prod_i 1_{\{ y_{i-1}y_i \in B_+\}}\\
&\text{and } \De(z,z') :=  1_{\{\|z\|_\infty \le \|z'\|_\infty\}} (\ftn(\|z\|_\infty) - \ftn(\|z'\|_\infty))^2 \text{ for } 
z,z' \in Y. 
\end{align*}
Since $\ftn'' \ge 0$ on $[n^{2/3},n]$ we can estimate 
$$
\De(z,z') \le \ftn'(\|z\|_\infty \vee n^{2/3})^2 \|z-z'\|_\infty^2 
\le \frac{9C^2}{\log n} \frac {\|z-z'\|_\infty^2} {(n^{2/3} \vee \|z\|_\infty)^2}1_{\La_n}(z)
$$
for $z,z' \in Y$. We thus obtain using \autoref{Lem:Ruellebound}
\begin{align*}
&\int  S_1' d\mu \le 
\int \!\! \mu (dY) \sumn_{y,y' \in Y} 2 \psi(y,y')\frac{9C^2}{\log n} 
\frac {\|y-y'\|_\infty^2}{(n^{2/3} \vee \|y\|_\infty)^2} 1_{\La_n}(y)\\
&\le \frac{18 \xi^2 C^2}{\log n} \int_{\La_n} \!\! dy \frac {1}{(n^{2/3} \vee \|y\|_\infty)^2} 
\int \! dy'  \psi(y,y') \|y-y'\|^2 
\end{align*}
and we can estimate the integrals using \eqref{equ:constants} and 
\begin{equation} \label{equ:intlog}
\int_{\La_n} \!\! dy \frac 1 {(n^{2/3} \vee \|y\|_\infty)^2}
=  \int_0^n \!\! ds \frac{8s}{(n^{2/3} \vee s)^2} = 4 + \frac 8 3 \log n \le 3 \log n
\end{equation} 
to get 
$$
\int S_1' d \mu  \le  54  \xi \de^4  c_\psi  \le \frac 1 {2\be} \de,
$$
provided that $\de$ is sufficiently small. Noting that $\|y_0-y_m\|_\infty \le m \|y_k-y_{k-1}\|_\infty$ for some $k$ we similarly 
obtain using \autoref{Lem:Ruellebound}
\begin{align*}
&\int S_1'' d\mu \otimes \pi_n \le 
\int \mu \otimes \pi_n(dY,dB)  \sum_{m \ge 1} \sum_{k=1}^m \quad \sumn_{y,y_0,...,y_m \in Y} \\
&\hspace*{4 cm} \psi(y,y_0)   \frac{18C^2}{\log n} \frac {m^2 \|y_k-y_{k-1}\|^2_\infty} {(n^{2/3} \vee \|y_0\|_\infty)^2}1_{\La_n}(y_0) \prod_i 1_{\{ y_{i-1}y_i \in B_+\}}\\
&\le \sum_{m \ge 1} \sum_{k=1}^m \int_{\La_n} \!\!\!\! dy_0 ... \int \! dy_m \int \! dy
\psi(y,y_0) \frac {18 C^2 m^2 \xi^2 \|y_k-y_{k-1}\|_\infty^2} {\log n(n^{2/3} \vee \|y_0\|_\infty)^2}\prod_i \xi g(y_{i-1},y_i),
\end{align*}
where we have used \eqref{equ:piint} and \autoref{Lem:Ruellebound} to estimate the integrals w.r.t. 
$\pi_n$ and $\mu$ and Tonelli's theorem to rearrange the order of sums and integrals. 
Next we estimate the integrals in the given order (starting with $\int dy$). 
Using \eqref{equ:constants}, \eqref{equ:intg} and \eqref{equ:intlog} 
we thus can estimate 
$$
\int \! S_1'' d\mu \otimes \pi_n 
\le 54 c_\psi \de^4  \xi c_u' \sum_{m \ge 1} m^3 c_{u}^{m-1} \le \frac 1 {2 \be} \de 
$$
if $\de$ is sufficiently small. This finishes the proof of 
Lemma \ref{Lem:energyestimate}. \qed 

\begin{Lem} For $\bY \in \bYY$ and $y,y' \in Y$ such that $(y,y') \in K^c$ we have 
$$
(y,y'+s (\tn(y')- \tn(y)) \ee) \in K^c \text{ for all } s \in [-1,1]. 
$$
\end{Lem}

\Pf This can be proved as in the proof of (5.9) in \cite{R2}. We include the argument 
for sake of completeness. W.l.o.g. $y := y_i \in C_i$ and $y' := y_j \in C_j$, where $0 \le i < j$. 
(For $i = j$ we have $\tn(y') = \tn(y)$ and are done.) Let 
$$
\La^i := \{y \in \R^2_S: t_i(y) \ge \tau_i\} \quad \text{ and } \quad 
K(y_i) :=  \{y \in \R^2_S: (y,y_i) \in K\}. 
$$
Furthermore let $T^s(y) := y + s t_{m+1}(y) \ee$. We observe that 
\begin{align*}
&T^s \text{ is continuous, $\le_{\ee}$-increasing and bijective, }\\
&T^s(\La^i) = \La^i + s\tau_i \ee \quad \text{ and } \quad T^s(\La_i \cap K(y_i)) = \La_i \cap K(y_i) + s\tau_i \ee.
\end{align*}
The first assertion follows exactly as \eqref{equ:Tcont}, the second follows from 
the first since we have $t_{m+1}(y) = \tau_i$ for all $y \in \partial \La^i$ (i.e. all $y$ s.t. 
$t_i(y) = \tau_i$) by definition of $t_{m+1}$, and the third follows from 
$t_{m+1}(y) = \tau_i$ for all $y \in K(y_i)\cap \La_i$ (since $t_{m+1}(y) \ge \tau_i$ for $y \in \La^i$
and $t_{m+1}(y) \le \tau_i$ for $y \in K(y_i)$). 
The three assertions together imply that 
$T^s(\La_i \stm K(y_i)) = \La_i \stm K(y_i) + s\tau_i \ee$. 
Since $y_j \in \La_i \stm K(y_i)$ we thus get $T^s(y_j) \in  \La_i \stm K(y_i) + s\tau_i \ee$, 
i.e. $y_j + s \tau_j \ee - s \tau_i\ee \notin K(y_i)$. \qed

\subsection{Density estimate: \autoref{Lem:densityestimate}}

We first note that by definition of $\th_n, \th_n^-$  we have 
$$
S_2(\bY) = \Big|\sum_{k \le m} \sum_{y \in P_k} \log |1- (\partial_{\ee} t_k(y))^2|\Big| 
\le 
\sum_{k \le m} \sum_{y \in P_k} \frac 4 3 (\partial_{\ee} t_k(y))^2, 
$$
where $P_k,t_k,m$ are from the construction of $\cTn(\bY)$. 
In the second step we have used 
$0 \ge \ln( 1 - a^2) \ge - \frac 4 3 a^2$ for $|a| \le \frac 1 2$ 
and $|\partial_{\ee} t_k(y)| \le \de \le \frac 1 2$ by
the $\de$-Lipschitz-continuity.
Now we would like to estimate $\partial_{\ee} t_k(y)$.  
In case of $y \in P_k$ with $k > m^*$ 
we have $\partial_{\ee} t_k(y) = 0$ (whenever the derivative exists), 
since (by definition of $m^*$) $t_{m^*+1}$
attains its maximal value $\tau_{m^*}$ at all points 
$y \in Y \stm \bigcup_{k \le m^*} C_k$. 
In case of $y \in P_k$ with $k \le m^*$ the derivative 
equals $\partial_{\ee} t_0(y)$ or $\partial_{\ee} \fm_{y',\tau_{j}}(y)$ for some 
$y' \in C_j$ such that  $j < k$ and $(y,y') \in K_\ep$. 
For the latter we have $\partial_{\ee} \fm_{y',\tau_{j}}(y) = 0$ in case of 
$(y,y') \in K$, and for $(y,y') \in K_{\ep} \stm K$ we have 
$$
|\partial_{\ee} \fm_{y',\tau_{j}}(y)| \le \frac{h_{y',\tau_j}}{\ep}
\le \frac {\ftn(\|y'\|_\infty - c_K) - \ftn(\|y''\|_\infty)} \ep \text{ for some } y'' \in C_{B_+}^\downarrow(y').
$$
Using $(a+b)^2 \le 2 a^2+2b^2$ we can thus estimate $S_2$ by $S_2' + S_2''+ S_2'''$, where 
\begin{align*}
&S_2'(Y) = \sum_{y \in Y} \frac 4 3 (\partial_{\ee} t_0(y))^2, \\
&S_2''(Y) = \sumn_{y,y' \in Y} 1_{K_\ep \stm K}(y,y') \frac 8 {3 \ep^2}
(\ftn(\|y'\|_\infty - c_K) - \ftn(\|y'\|_\infty)^2 \text{ and } \\
&S_2'''(Y,B) = \sum_{m \ge 1} \quad \sumn_{y,y_0,...,y_m \in Y}  \frac 8 {3\ep^2} 
 1_{K_\ep \stm K}(y,y_0) \De(y_0,y_m) \prod_i 1_{\{y_iy_{i-1} \in B_+\}}.
\end{align*}
Proceeding as in the estimates of the last subsection we obtain 
\begin{align*}
&\int \! S_2' d\mu \le \int \!\! dy   \frac 4 3 \frac{9C^2\xi}{\log n} \frac 1 {(n^{2/3} \vee (\|y\|_\infty \wedge n))^2}  
= \frac{12C^2 \xi}{\log n}  \int_{n^{2/3}}^n \!\!\!\! ds \frac{8s}{s^2} \le 32 \de^4 \xi \le \frac{\de}{3}
\end{align*}
provided that $\de$ is sufficiently small. For the second contribution we have 
\begin{align*}
&\int S_2'' d\mu \le \int \! dy' \frac {8\xi} {3 \ep^2}
(\ftn(\|y'\|_\infty - c_K) - \ftn(\|y'\|_\infty)^2 \int \! dy 1_{K_\ep \stm K^U}(y,y') \xi 
\end{align*} 
The last integral can be estimated by $1$ using \eqref{equ:constants}.  
For the remaining integral we note that
\begin{align*}
&\int (\ftn(\|y'\|_\infty - c_K) - \ftn(\|y'\|_\infty)^2  dy'
= \int_{n^{2/3}}^{n+c_K} 8s (\ftn(s - c_K) - \ftn(s))^2 ds \\
&\le \int_{1+c_K}^{n+c_K} 9(s-c_K) c_K^2 \frac{9C^2}{\log n} \frac 1 {(s-c_K)^2} ds 
= 81 C^2c_K^2.  
\end{align*}
In the second step we have used that $n^{2/3} \ge 1+c_K$ and $8s \le 9(s-c_K)$ for $s \ge n^{2/3}$, which is due to 
$\de < \frac 1 {c_K}$ and $n \ge \frac 1 {\de^8}$. We thus get 
$$
\int S_2'' d\mu \le \frac{216\xi}{\ep^2} \de^4 c_K^2
\le \frac{\de} {3}
$$
if $\de$ is sufficiently small. 
For $S_2'''$ we use the estimate of $\De$ from the last subsection together with 
$\|y_0-y_m\|_\infty \le m \|y_k-y_{k-1}\|_\infty$ for some $k$. Proceeding as in the estimates of the last subsection we obtain  
\begin{align*}
&\int S_2''' d \mu \otimes \pi_n
\le  \sum_{m \ge 1} \sum_{k=1}^m\int_{\La_n} dy_0 .... \int dy_m \int dy \\
&\hspace*{2 cm}  \frac 8 {3\ep^2} \frac{9C^2 \xi}{\log n} \frac{m^2\|y_k-y_{k-1}\|^2}{(n^{2/3} \vee \|y_0\|_\infty)^2} g(y_0,y) \xi  \prod_i \xi g(y_{i-1},y_i)\\
&\le \sum_{m \ge 1} \sum_{k=1}^m \frac{24 C^2 \xi}{\ep^2} 3 m^2 c_{u}^m c_u'
\le \frac{72 \de^4 \xi c_u'}{\ep^2} \sum_{m \ge 1} m^3 c_{u}^m \le \frac{\de} {3}
\end{align*}
for $\de$ sufficiently small. 
This finishes the proof of Lemma \eqref{Lem:densityestimate}. \qed


\begin{thebibliography}{FP1}


\bibitem[BK]{BK}
E. P. Bernard, W. Krauth, Two-step melting in two dimensions: First-order liquid-hexatic transition,  Phys. rev. letters 107(15) (2011), 155704.

\bibitem[Eea]{Eea}

M. Engel, J. A. Anderson, S. C. Glotzer, E. P. Bernard, W. Krauth, Hard-disk equation of state: 
First-order liquid-hexatic transition in two dimensions with three simulation methods, 
Phys. rev. E 87(4) (2013), 042134.


\bibitem[FP1]{FP1}
J. Fr\"ohlich, C.-E. Pfister, On the absence of spontaneous symmetry
breaking and of crystalline ordering in two-dimensional systems,
Comm. Math. Phys. 81 (1981) 277-298.

\bibitem[FP2]{FP2}
J. Fr\"ohlich, C.-E. Pfister, Absence of crystalline ordering 
in two dimensions, Comm. Math. Phys. 104 (1986) 697-700.


\bibitem[KK]{KK}
S. C. Kapfer, W. Krauth, Two-dimensional melting: From liquid-hexatic coexistence to continuous transitions, Phys. rev. letters 114(3) (2015), 035702.


\bibitem[M]{M}
N. D. Mermin, Absence of ordering in certain classical systems,
J. Math. Phys. 8 (1967) 1061-1064.

\bibitem[MP]{MP}
P. Mi\l o\'{s}, R. Peled, 
Delocalization of two-dimensional random surfaces with hard-core constraints,
Comm. Math. Phys. 340(1) (2015) 1-46. 


\bibitem[MW]{MW}
N. D. Mermin, H. Wagner, Absence of ferromagnetism or
antiferromagnetism in one- or two-dimensional isotropic Heisenberg
models, Phys. Rev. Letters 17 (1966) 1133-1136.

\bibitem[P1]{P1}
R. Peierls, Remarks on transition temperatures, Helv. Phys. Acta, 
7(2) (1934) 81-83.

\bibitem[P2]{P2}
R. Peierls, Quelques propri\'et\'es typiques des corps solides,  Ann. de l'Institut Henri Poincar\'e 5 (3) (1935) 177-222.

\bibitem[Ri1]{R1}
T. Richthammer, Translation-invariance of two-dimensional Gibbsian point processes,  Comm. Math. Phys. 274(1) (2007) 81-122.

\bibitem[Ri2]{R2}
T. Richthammer,  Translation invariance of two-dimensional Gibbsian systems 
of particles with internal degrees of freedom, Stoch. Proc. Appl.  119(3) (2009) 700-736.

\bibitem[Ri3]{R3}
T. Richthammer, Lower Bound on the Mean Square Displacement of Particles in the Hard Disk Model,
Comm. Math. Phys. 345(3)  (2016) 1–23.

\bibitem[Ru]{Ru} D. Ruelle, Superstable interactions in classical statistical mechanics, Comm. Math. Phys. 18 (1970) 127–159. 


\bibitem[WR]{WR}
B. Widom, J. S. Rowlison, 
New Model for the Study of Liquid-Vapor Phase Transitions, 
J. Chem. Phys., 52 (4) (1970) 1670–1684.
\end{thebibliography}
\end{document}